\documentclass[12pt,a4paper]{article}
\usepackage[english]{babel}
\usepackage[letterpaper, top=0.95in, bottom=0.95in, left=0.95in, right=0.95in, marginparwidth=1.7cm]{geometry}
\usepackage{xcolor}
\definecolor{darkblue}{rgb}{0,0,.6}
\usepackage{amsmath}
\usepackage{graphicx}
\usepackage{amsfonts, microtype}
\usepackage{placeins}
\usepackage{subcaption}
\usepackage{booktabs,setspace}
\usepackage{footnote,longtable}
\makesavenoteenv{tabular}
\usepackage[round]{natbib}
\usepackage[colorlinks=true, allcolors=darkblue]{hyperref}
\usepackage{bbm,url}
\usepackage{multirow}
\usepackage{amssymb, orcidlink}
\usepackage{soul}
\graphicspath{{Figures/}}
\newcommand{\X}{\mathcal{X}}

\usepackage[dvipsnames]{xcolor}

\title{On the Distributed Estimation for Scalar-on-Function Regression Models}
\author{\normalsize Peilun He \orcidlink{0000-0002-2740-3390}\\
\normalsize Department of Statistics and Applied Probability \\
\normalsize University of California, Santa Barbara \\
\normalsize Department of Actuarial Studies and Business Analytics \\
\normalsize Macquarie University \\
\\
\normalsize Han Lin Shang \orcidlink{0000-0003-1769-6430}\\
\normalsize Department of Actuarial Studies and Business Analytics \\
\normalsize Macquarie University \\
\\
\normalsize Nan Zou \orcidlink{0000-0001-6907-1294} \\
\normalsize School of Mathematical and Physical Sciences \\
\normalsize Macquarie University}
\date{}

\begin{document}
\maketitle

\begin{abstract}
\{This paper provides a distributed estimation framework for point and interval inference in scalar-on-function regression models.\}
    
This paper proposes distributed estimation procedures for three scalar-on-function regression models: the functional linear model (FLM), the functional non-parametric model (FNPM), and the functional partial linear model (FPLM). The framework addresses two key challenges in functional data analysis, namely the high computational cost of large samples and limitations on sharing raw data across institutions. Monte Carlo simulations show that the distributed estimators substantially reduce computation time while preserving high estimation and prediction accuracy for all three models. When block sizes become too small, the FPLM exhibits overfitting, leading to narrower prediction intervals and reduced empirical coverage probability. An example of an empirical study using the \textit{tecator} dataset further supports these findings.

\vspace{0.5em}

\end{abstract}


\newpage

\setstretch{1.29}

\section{Introduction}
\label{sec:intro}

Regression with functional data is one of the most thoroughly researched topics within the broader literature on functional data analysis. Regression models can be categorized into three groups by the role played by the functional data in each model: scalar-valued response and function-valued predictor (``scalar-on-function'' regression); function-valued response and scalar-valued predictor (``function-on-scalar'' regression); and function-valued response and predictor (``function-on-function'' regression). This paper focuses on the first case and revisit a linear model, a partial linear model, a non-parametric model to scalar-on-function regression. Domains where scalar-on-function regression has been applied include chemometrics \citep{goutis98}, cardiology \citep{RHL02b, RLH02}, brain science \citep{RO10}, climate science \citep{FLV05} and many others. For a comprehensive review, refer to \cite{morris2015}, \cite{RGS+17} and \cite{KS23}. 

Historically, scalar-on-function regression has typically been conducted on relatively small datasets using a single computing machine. In contrast, modern applications increasingly involve functional data with massive sample sizes that are distributed across multiple databases, such as those maintained by financial or healthcare institutions. As sample sizes grow, the computational cost of fitting functional regression models becomes substantial. Moreover, in many applications, raw data are sensitive and cannot be transferred to a central server due to privacy or regulatory constraints. For example, patient medical records are protected by privacy regulations and cannot be shared across hospitals or research institutions. Similarly, customer-level transaction records held by banks are highly confidential and cannot be shared with other financial institutions. These challenges make it impractical to aggregate all data on a single machine for analysis.

A common solution is to perform computations locally and then aggregate intermediate results at a central server. To address such distributed data settings, distributed learning approaches have been developed to efficiently communicate and summarise information from local databases. For a review of distributed learning methods for generic data, see \cite{mcdonald2009efficient}, \cite{li2013statistical}, \cite{zhang2015divide}, \cite{chang2017divide}, \cite{zhou2024distributed}, \cite{xiao2024distributed}. In the context of functional data, theoretical developments have been provided by \cite{tong2021distributed}, \cite{cai2024optimal}, \cite{xue2024optimal}, \cite{liu2024distributed}, \cite{liu2024statistical}, \cite{liu2025distributed}. This manuscript focuses on the empirical performance of distributed learning methods for scalar-on-function regression.

This paper proposes a distributed methodology for point estimation, point prediction, and prediction interval construction in three scalar-on-function models: the functional linear model, the functional non-parametric model, and the functional partial linear model. The proposed approach allows each model to be fitted locally on partitioned subsets of the data, with only aggregated intermediate results transmitted to a central server, thereby reducing computational burden and avoiding the need to share raw, sensitive data. The computational cost and statistical accuracy of the methodology are examined through extensive Monte Carlo simulations and an application to the tecator dataset. The results demonstrate that the distributed method achieves substantial reductions in execution time while maintaining high prediction accuracy and preserving data privacy. Although divide-and-conquer estimators have been widely studied in functional data analysis, existing work has largely focused on the linear setting \citep[see, e.g.,][]{tong2021distributed, xue2024optimal, liu2025distributed}. In contrast, this paper considers linear, non-parametric, and partial linear scalar-on-function models within a unified distributed framework.

This paper is structured as follows. Section~\ref{sec:model} defines three models, the scalar-on-function linear model, scalar-on-function non-parametric model, and scalar-on-function partial linear model, and describes the corresponding parameter estimation methods. Section~\ref{sec:method} describes the distributed methods for point estimation and interval estimation. Sections~\ref{sec:simulation} and~\ref{sec:real_data} present the results of the simulation study and the \textit{tecator} data analysis, respectively. Finally, Section~\ref{sec:conclusion} concludes, along with some ideas on how the methodology can be extended.

\section{Scalar-on-Function Regression Models}\label{sec:model}

We consider three types of scalar-on-function regression models: the functional linear model (FLM), functional non-parametric model (FNPM), and functional partial linear model (FPLM). Particularly, for FLM, both B-spline expansion and functional principal component analysis (FPCA) are employed as estimation methods. For FNPM and FPLM, we adopt a kernel-based approach that extends the Nadaraya–Watson estimator to the functional setting.

\subsection{Scalar-on-Function Linear Model}\label{sec:FLM}

We begin with a scalar-on-function linear regression model, defined as
\begin{equation}
Y_{i} = \int_{\mathcal{T}} \beta(t) \mathcal{X}_i(t) dt + \epsilon_{i}, 
\label{eq:FLM}
\end{equation}
where $Y_i \in \mathbb{R}$ denotes a scalar-valued response for $i = 1, \dots, N$, $\mathcal{X}_i \in \mathbb{H}$ is a functional covariate taking random values in a Hilbert space $\mathbb{H}$ with domain $\mathcal{T}$, $\beta \in \mathbb{H}$ is the corresponding functional coefficient, and $\epsilon_i \in \mathbb{R}$ is an random error term independent of $\mathcal{X}_i$, satisfying $\mathbb{E}(\epsilon_i) = 0$ and $Var(\epsilon_i) = \sigma^2 < \infty$. 

Each function $\mathcal{X}_i(\cdot)$ is assumed to be observed only at discrete points $t_1, \dots, t_M$. The objective of the parametric functional regression model is to estimate $\beta$ by finding $\widehat{\beta}$ that minimises the quadratic loss function
\begin{equation*}
L(\widehat{\beta}) = \frac{1}{N} \sum_{i=1}^N \left( Y_i - \int_{\mathcal{T}} \widehat{\beta}(t) \mathcal{X}_i(t)dt \right)^2.
\end{equation*}

\cite{ramsay2005functional} employs a basis expansion approach to estimate the functional coefficient $\beta(\cdot)$, thereby reducing the infinite-dimensional covariate to a finite-dimensional representation. The functional covariate $\mathcal{X}_i(t)$ is approximated as
\begin{equation*}
    \mathcal{X}_i(t) \approx \sum_{p=1}^P c_{ip} \phi_p(t),
\end{equation*}
where $\phi_p(t)$ denotes known orthogonal or non-orthogonal basis functions, and $c_{ip}$ are the associated coefficients. At discrete time points $t_1, \dots, t_M$, we observe $\mathcal{X}_i(t_1), \dots, \mathcal{X}_i(t_M)$ and $\phi_p(t_1), \dots, \phi_p(t_M)$. The coefficients $c_{ip}$ are estimated by minimising the least-squares criterion
\begin{equation*}
    \sum_{j=1}^M \left[ \mathcal{X}_i(t_j) - \sum_{p=1}^P c_{ip} \phi_p(t_j) \right]^2. 
\end{equation*}

Similarly, the functional coefficient $\beta(t)$ is represented as
\begin{equation*}
    \beta(t) \approx \sum_{q=1}^Q b_q \psi_q(t),
\end{equation*}
where $\psi_q(t)$denotes another set of basis functions with coefficients $b_q$. Without loss of generality, $\mathcal{X}_i(t)$ and $\beta(t)$ may employ different basis sets $\phi_p(t)$ and $\psi_q(t)$, although they can also share the same basis. Further, we assume that both $P$ and $Q$ are fixed and known. 

Substituting these expansions into~\eqref{eq:FLM} yields
\begin{align}
Y_i &= \int_{\mathcal{T}} \beta(t) \mathcal{X}_i(t) dt + \epsilon_i \nonumber \\
&= \int_{\mathcal{T}} \sum_{q=1}^Q b_{q} \psi_q(t) \sum_{p=1}^{P} c_{ip} \phi_{p}(t) dt + \epsilon_{i} \nonumber \\ 
&= \sum_{q=1}^{Q} b_{q} \int_{\mathcal{T}} \psi_{q}(t) \sum_{p=1}^{P} c_{ip} \phi_{p}(t) dt + \epsilon_{i} \nonumber \\
&= \sum_{q=1}^{Q} b_{q} U_{iq} + \epsilon_{i}, \nonumber
\end{align}
which corresponds to a standard linear regression model, where
\begin{equation*}
    U_{iq} = \int_{\mathcal{T}} \psi_q(t) \sum_{p=1}^P c_{ip} \phi_p(t) dt
\end{equation*}
is deterministic. The coefficients $b_q$ are then estimated by minimising
\begin{equation*}
    \sum_{i=1}^{N} \left[Y_{i} - \sum_{q=1}^{Q} b_{q} U_{iq} \right]^2,
\end{equation*}
and the estimated functional coefficient is obtained as
\begin{equation*}
    \widehat{\beta}(t) = \sum_{q=1}^{Q} \widehat{b}_q \psi_q(t).
\end{equation*}

We consider two choices of basis functions $\phi_p(t)$ and $\psi_q(t)$. The first choice is the B-spline basis, where both $\phi_p(t)$ and $\psi_q(t)$ are B-spline functions, possibly with different numbers of basis elements. In this case, the basis functions are non-orthogonal and deterministic. The second choice employs eigenfunctions obtained via functional principal component analysis (FPCA) on the empirical covariance function of $\mathcal{X}_i(t)$. Here, $\phi_p(t)$ and $\psi_q(t)$ are the orthonormal eigenfunctions, $Q = P$, and the basis is data-driven. We refer to these two approaches as the B-spline expansion and the FPCA method, respectively.

\subsection{Scalar-on-Function Non-Parametric Model}
\label{sec:FNPM}

The functional non-parametric model (FNPM) is defined as
\begin{equation}
    Y_i = m(\mathcal{X}_i(t)) + \epsilon_i,
    \label{eq:FNPM}
\end{equation}
where $m: \mathbb{H} \to \mathbb{R}$ is an unknown, possibly non-linear function. The error term $\epsilon_i$ is assumed to be independent of $\mathcal{X}_i$, with $\mathbb{E}(\epsilon_i) = 0$ and $Var(\epsilon_i) = \sigma^2 < \infty$. This formulation allows complete flexibility in modelling the relationship between the functional covariate and the scalar response, without imposing any specific parametric structure.

Estimating the unknown function $m(\cdot)$ is a key challenge in the FNPM, primarily due to the non-linearity of $m(\cdot)$, which makes the linear estimation techniques in Section~\ref{sec:FLM} inapplicable. We adopt a kernel-based approach following \citet{ferraty2006nonparametric}, which extends the Nadaraya–Watson estimator to the functional setting. The kernel estimator of $m(\cdot)$ is given by
\begin{equation}
\widehat{m}(\mathcal{X}) = \frac{\sum_{i=1}^N Y_i K(h^{-1} d(\mathcal{X}, \mathcal{X}_i))}{\sum_{i=1}^N K(h^{-1} d(\mathcal{X}, \mathcal{X}_i))},
\label{eq:est_FNPM}
\end{equation}
where $K: \mathbb{R} \to \mathbb{R}$ is an asymmetric kernel function, $h > 0$ is the bandwidth parameter, and $d: \mathbb{H} \times \mathbb{H} \to \mathbb{R}$ is a semi-metric measuring the proximity between functional observations.

Defining the normalised kernel weights as
\begin{equation*}
w_h(\mathcal{X}, \mathcal{X}_i) = \frac{K(h^{-1} d(\mathcal{X}, \mathcal{X}_i))}{\sum_{i=1}^N K(h^{-1} d(\mathcal{X}, \mathcal{X}_i))},
\end{equation*}
the estimator~\eqref{eq:est_FNPM} can be rewritten as
\begin{equation*}
    \widehat{m}(\mathcal{X}) = \sum_{i=1}^N w_h(\mathcal{X}, \mathcal{X}_i) Y_i, 
\end{equation*}
which represents a weighted average of the response variables $Y_i$, where the weights satisfy $\sum_{i=1}^{N} w_{h}(\mathcal{X}, \mathcal{X}_i)$ =~1. The asymptotic properties of this estimator have been extensively investigated in the literature \citep[see, e.g.,][]{masry2005nonparametric, ferraty2006nonparametric, ferraty2007nonparametric}.

We employ an asymmetric normal kernel defined as
\begin{equation*}
K(x) = \begin{cases} 
\frac{2}{\sqrt{2\pi}} e^{-x^2/2} \quad \text{if} \; x \ge 0 \\ 
0 \quad \quad \quad \quad \;\;\; \text{if} \; x < 0 
\end{cases} 
\end{equation*}
and the semi-metric between two functional observations is given by
\begin{equation}
    d(\mathcal{X}, \mathcal{X}_i) = \left\Vert \mathcal{X} - \mathcal{X}_i \right\Vert_2 = \left( \int_{\mathcal{T}} \left\vert \mathcal{X}(t) - \mathcal{X}_i(t) \right\vert^2 dt \right)^{1/2}.
    \label{eq:dist}
\end{equation}
The integral is approximated using Simpson’s rule, and the bandwidth parameter $h$ is selected via cross-validation.

\subsection{Scalar-on-Function Partial Linear Model}
\label{sec:FPLM}

The functional partial linear model (FPLM) extends the FNPM by incorporating a linear relationship between the scalar response $Y_i$ and an additional scalar covariate $Z_i$:
\begin{equation}
Y_{i} = \beta_{\text{NF}} Z_{i} + m(\mathcal{X}_i(t)) + \epsilon_{i},
\label{eq:FPLM}
\end{equation}
where $Z_{i} \in \mathbb{R}$ denotes a non-functional covariate with associated coefficient $\beta_{\text{NF}} \in \mathbb{R}$. Throughout this paper, the subscript ``NF'' is used to distinguish coefficients associated with non-functional covariates from those linked to functional predictors. The objective in this partial linear setting is to jointly estimate both $\beta_{NF}$ and the non-linear functional component $m(\cdot)$.

Unlike the FNPM, this formulation includes an additional linear term, $\beta_{NF} Z_i$, which complicates estimation. In this case, neither the kernel estimator nor ordinary least squares can be applied directly.

Following \cite{aneiros2006semi}, we define $\boldsymbol{Z} = [Z_1, \dots, Z_N]^\top$, $\boldsymbol{Y} = [Y_1, \dots, Y_N]^\top$, and $\widetilde{\boldsymbol{Z}}_h = (\boldsymbol{I} - \boldsymbol{W}_h) \boldsymbol{Z}$, $\widetilde{\boldsymbol{Y}}_h = (\boldsymbol{I} - \boldsymbol{W}_h) \boldsymbol{Y}$, where $\boldsymbol{W}_h = \{ w_h(\mathcal{X}_i, \mathcal{X}_j)\}_{N \times N}$ is the matrix of kernel weights. Then, $\beta_{NF}$ is estimated as
\begin{equation}
\widehat{\beta}_{NF} = (\widetilde{\boldsymbol{Z}}_h^\top \widetilde{\boldsymbol{Z}}_h)^{-1} \widetilde{\boldsymbol{Z}}_h^\top \widetilde{\boldsymbol{Y}}_h \label{eq:beta_est_partial}
\end{equation}
and the functional component $m(\cdot)$ is estimated as
\begin{equation}
\widehat{m}(\mathcal{X}) = \sum_{i=1}^N w_h(\mathcal{X}, \mathcal{X}_i) (Y_i - \widehat{\beta}_{NF} Z_i) = \frac{\sum_{i=1}^N K(h^{-1} d(\mathcal{X}, \mathcal{X}_i)) (Y_i - \widehat{\beta}_{NF} Z_i)}{\sum_{i=1}^N K(h^{-1} d(\mathcal{X}, \mathcal{X}_i))}.
\label{eq:est_partial}
\end{equation}
The fitted response is then given by
\begin{equation*}
    \widehat{Y}_i = \widehat{\beta}_{NF} Z_i + \widehat{m}(\mathcal{X}_i). 
    \label{eq:Y_hat_FPLM}
\end{equation*}

These partial linear estimators have been further extended in various directions. For instance, \citet{novo2021knn} proposed a fast and flexible $k$-nearest-neighbour (kNN) extension of the kernel estimator to improve local adaptivity,
while \citet{novo2021sparse} introduced a penalised version of~\eqref{eq:beta_est_partial} to enable covariate selection.

For the FPLM, the choices of the kernel function $K(\cdot)$ and the semi-metric $d(\cdot, \cdot)$ are the same as those used in the FNPM. 

\section{Methodologies}\label{sec:method}

Although Section~\ref{sec:model} introduces estimation methods for functional regression models, these approaches may face practical limitations. First, as the sample size increases, computational time can grow substantially. 
Second, in many applications, raw data cannot be shared between institutions due to privacy or regulatory constraints. For example, hospitals may each collect data from their own patients but are prohibited from sharing these records externally. Therefore, it is essential to develop methodologies that enable model estimation to be performed locally on each site’s data, with the resulting estimates subsequently aggregated on a central server. In Section~\ref{sec:point_est}, we propose distributed estimation procedures for the functional regression models~\eqref{eq:FLM},~\eqref{eq:FNPM}, and~\eqref{eq:FPLM}. These distributed estimators substantially reduce computational cost while preserving data privacy. Section~\ref{sec:intercal_est} introduces the distributed procedures for interval estimation and prediction accuracy. 

\subsection{Distributed Point Estimation}
\label{sec:point_est}

Let $\mathcal{D} = \{ (\mathcal{X}_i, Y_i) \}_{i=1}^{N_{\mathrm{train}}}$ denote the training dataset of size $N_{\mathrm{train}}$ and let $S = \{1, \dots, N_{\text{train}}\}$ denote the index set of the training samples. The dataset is partitioned into $K$ non-overlapping subsets $S_k \subset S$, where $S_k$ corresponds to the indices of the $k$\textsuperscript{th} block.\footnote{Only the training data are partitioned into blocks, whereas the testing data are used in full for all experiments.} This partition satisfies $\cup_{k=1}^K S_k = S$ and $S_{k_1} \cap S_{k_2} = \varnothing$ for any $k_1 \ne k_2$. For simplicity, we assume all blocks are of equal size, i.e. $|S_k| = \frac{N_{\mathrm{train}}}{K} = n$ for all $k = 1, \dots, K$. 

A local estimator is computed within each block. Let $\widehat{Y}^{(k)}(\cdot)$ denote the local estimator of the response variable in the $k$\textsuperscript{th} block. Since the blocks are non-overlapping, the global (distributed) estimator for the training data can be expressed as
\begin{equation*}
    \widehat{Y}(\mathcal{X}_i) = \sum_{k=1}^K \widehat{Y}^{(k)}(\mathcal{X}_i) \cdot \mathbbm{1} \left[ i \in S_k \right],
\end{equation*} 
where $\mathbbm{1}[\cdot]$ denotes the indicator function. 

For the FLM, we also estimate the functional coefficient $\beta$. Let $\widehat{\beta}^{(k)}$ denote the local estimator obtained from the $k$\textsuperscript{th} block. The global estimator is then obtained by averaging across all local estimates:
\begin{equation}
    \widehat{\beta} = \frac{1}{K} \sum_{k=1}^K \widehat{\beta}^{(k)}. 
    \label{eq:dist_beta}
\end{equation}

We now turn to prediction for new observations. Let $\mathcal{D}^* = \{ (\mathcal{X}_\iota^*, Y_\iota^*) \}_{\iota=1}^{N_{\mathrm{test}}}$ denote the testing dataset of size $N_{\mathrm{test}}$. For any new functional covariate $\mathcal{X}_{\iota}^*$, each local estimator $\widehat{Y}^{(k)}(\cdot)$ produces a local prediction, and the final global prediction is obtained by averaging across all $K$ blocks:
\begin{equation}
    \widehat{Y}(\mathcal{X}_{\iota}^*) = \frac{1}{K} \sum_{k=1}^K \widehat{Y}^{(k)}(\mathcal{X}_{\iota}^*).
    \label{eq:dist_test}
\end{equation}


\subsection{Distributed Interval Estimation}
\label{sec:intercal_est}

We present a distributed procedure for constructing prediction intervals for the response variable $Y_i$ using a standard-deviation-based method originally proposed by \citet{shang2025constructing}.

Let $Y_{i}^{(k)}$ and $\widehat{Y}_{i}^{(k)}$ denote the observed and fitted values of the $i$\textsuperscript{th} observation in the $k$\textsuperscript{th} block, respectively. The estimation error is defined as
\begin{equation*}
    \widehat{\epsilon}_i^{(k)} = Y_i^{(k)} - \widehat{Y}_i^{(k)}.
\end{equation*}
The standard deviation of the residuals within this block is then computed as
\begin{equation*}
    \widehat{\sigma}^{(k)} = \textrm{sd} \left( \widehat{\epsilon}_i^{(k)} \right).
\end{equation*}

For a given significance level $\alpha$, we seek a value $\gamma^{(k)}$ such that
\begin{equation*}
    \mathrm{Pr} \left( - \gamma^{(k)} \widehat{\sigma}^{(k)} \le \widehat{\epsilon}_i^{(k)} \le \gamma^{(k)} \widehat{\sigma}^{(k)} \right)
\end{equation*}
approximates the nominal coverage probability $(1-\alpha) \times 100\%$. By the law of large numbers, this probability can be approximated empirically as
\begin{equation*}
    \mathrm{Pr} \left( - \gamma^{(k)} \widehat{\sigma}^{(k)} \le \widehat{\epsilon}_i^{(k)} \le \gamma^{(k)} \widehat{\sigma}^{(k)} \right) \approx \frac{1}{n} \sum_{i=1}^{n} \mathbbm{1} \left[ - \gamma^{(k)} \widehat{\sigma}^{(k)} \le \widehat{\epsilon}_i^{(k)} \le \gamma^{(k)} \widehat{\sigma}^{(k)} \right].
\end{equation*}
The quantity $\gamma^{(k)}$ is obtained by minimising the coverage probability difference (CPD):
\begin{equation*}
    \textrm{CPD}^{(k)} = \left| \frac{1}{n} \sum_{i=1}^{n} \mathbbm{1} \left[ - \gamma^{(k)} \widehat{\sigma}^{(k)} \le \widehat{\epsilon}_i^{(k)} \le \gamma^{(k)} \widehat{\sigma}^{(k)} \right] - (1-\alpha) \right|.
\end{equation*}
The global estimates $\gamma$ and $\widehat{\sigma}$ are then obtained by averaging across the $K$ blocks:
\begin{align}
    \gamma = \frac{1}{K} \sum_{k=1}^K \gamma^{(k)}, \label{eq:dist_gamma}\\
    \widehat{\sigma} = \frac{1}{K} \sum_{k=1}^{K}\widehat{\sigma}^{(k)}. \label{eq:dist_sd}
\end{align}
For the testing data, the same global values of $\gamma$ and $\widehat{\sigma}$ are employed. The prediction interval is 
\begin{equation*}
    \widehat{Y}_{\iota}^* \pm \gamma \widehat{\sigma}, 
\end{equation*}
where $\widehat{Y}_{\iota}^*$ is the global prediction obtained from the distributed point estimation procedure described in Section~\ref{sec:point_est}. 

\section{Monte Carlo Simulation}\label{sec:simulation}

This section presents the results of the simulation study. The data-generating processes (DGPs) are described in Section~\ref{sec:sim_DGP}, where the training and testing datasets are independently generated for each model. The evaluation criteria used to assess both in-sample estimation and out-of-sample prediction accuracy are outlined in Section~\ref{sec:sim_acc}. Section~\ref{sec:sim_results} reports the results for the FLM, FNPM, and FPLM, respectively.

\subsection{Data Generating Process}
\label{sec:sim_DGP}
\paragraph{DGP for the FLM.}
For the FLM in \eqref{eq:FLM}, we use the same DGP as in \cite{beyaztas2022robust} and \cite{beyaztas2024robust}. For any $t \in [0, 1]$, the functional covariate is generated as 
\begin{equation*}
    \mathcal{X}_i(t) = \sum_{j=1}^5 k_{ij} v_j(t),
\end{equation*}
where $k_{ij} \sim N(0, 4j^{-3/2})$. The basis functions are defined as 
\begin{equation*}
    v_j(t) = \sin{(j \pi t)} - \cos{(j \pi t)},
\end{equation*} 
and the coefficient function $\beta(t)$ is specified as 
\begin{equation*}
    \beta(t) = 2 \sin{(2 \pi t)}.
\end{equation*}
The response variable $Y_i$ is generated according to 
\begin{equation*}
    Y_i = \int_0^1 \beta(t) \mathcal{X}_i(t) dt + \epsilon_i,
\end{equation*}
where $\epsilon_i \sim N(0, 1)$. In the discrete setting, the domain $[0, 1]$ is observed on a equally spaced grid with step size 0.01. 

\paragraph{DGP for the FNPM.} 
For the FNPM in \eqref{eq:FNPM}, we employ the DGP proposed by \cite{ferraty2007nonparametric}. For any $t \in [-1, 1]$, the functional covariate is generated as 
\begin{equation*}
\mathcal{X}_i(t) = \cos{(\omega_i t)} + (a_i + 2 \pi) t + b_i,
\end{equation*}
where $a_i$ and $b_i$ are uniformly distributed on $[0,1]$, and $\omega_i$ is uniformly distributed on $[0, 2\pi]$. The function $m(\cdot)$ is given by
\begin{equation*}
    m(\mathcal{X}(t)) = \int_{-1}^{1} |\mathcal{X}'(t)| (1 - \cos{(\pi t)}) dt,
\end{equation*}
and the error term $\epsilon_i$ follows $N(0,2)$.

\paragraph{DGP for the FPLM.}
For the FPLM in \eqref{eq:FPLM}, we adopt the same simulation setup for $\mathcal{X}_i(t)$, $Y_i$, and $m(\cdot)$ as in the FNPM. The non-functional covariate $Z_i$ is drawn from a standard normal distribution, and the non-functional coefficient is set to $\beta_{NF} = 0.5$. 

\subsection{Evaluation Criteria}
\label{sec:sim_acc}
We now introduce the criteria used to evaluate model performance. To assess the \textit{in-sample} estimation accuracy of the response variable $Y_i$ for the training data, we use the root mean squared error (RMSE), relative error (RE), and mean absolute error (MAE), defined as
\begin{align*}
\mathrm{RMSE} &= \sqrt{\frac{1}{N_{\text{train}}}\sum_{k=1}^K \sum_{i = 1}^n \left( Y_i^{(k)} - \widehat{Y}_i^{(k)} \right)^2}, \\
\mathrm{RE} &= \frac{1}{N_{\mathrm{train}}} \sum_{k=1}^K \sum_{i=1}^n \left| \frac{Y_i^{(k)} - \widehat{Y}_i^{(k)}}{Y_i^{(k)}} \right|, \\
\mathrm{MAE} &= \frac{1}{N_{\mathrm{train}}} \sum_{k=1}^K \sum_{i=1}^n \left| Y_i^{(k)} - \widehat{Y}_i^{(k)} \right|,  
\end{align*}
where 
$n = N_{\mathrm{train}} / K$ is the number of observations in each block, and $Y_i^{(k)}$ and $\widehat{Y}_i^{(k)}$ denote the true and estimated responses for the $i$\textsuperscript{th} observation in the $k$\textsuperscript{th} block.

For the FLM, we also assess the estimation accuracy of the functional coefficient $\beta(t)$ using the rescaled Frobenius norm, the pointwise bias, and the pointwise standard deviation. The rescaled Frobenius norm (F.Norm) is defined as
\begin{equation*}
\mathrm{F.Norm} = \left\Vert \boldsymbol{\beta} - \boldsymbol{\widehat{\beta}} \right\Vert_F^2 \approx 
\frac{1}{M}\sum_{m=1}^{M} \left( \beta(t_m) - \widehat{\beta}(t_m) \right)^2, 
\end{equation*}
where $\beta(\cdot)$ is evaluated on a dense grid of $M$ points $t_1, \dots, t_M$, and 
\begin{equation*}
    \boldsymbol{\beta} = [\beta(t_1), \dots, \beta(t_M)]^\top, \qquad \boldsymbol{\widehat{\beta}} = [\widehat{\beta}(t_1), \dots, \widehat{\beta}(t_M)]^\top.
\end{equation*}
The global estimator $\boldsymbol{\widehat{\beta}}$ is obtained using~\eqref{eq:dist_beta}. 

The pointwise bias and pointwise standard deviation (ST.DEV) across Monte Carlo replications are computed as
\begin{align*}
\mathrm{Bias^{2}} &= \frac{1}{M} \sum_{m=1}^M \left[ \mathbb{E}\left( \widehat{\beta}(t_m) \right) - \beta(t_m) \right]^2, \\
\mathrm{ST.DEV} &= \sqrt{\frac{1}{M} \sum_{m=1}^M \mathrm{Var}\left( \widehat{\beta}(t_m) \right)}.
\end{align*}

To evaluate \textit{out-of-sample} prediction accuracy for the testing data, we use the root mean square forecast error (RMSFE), relative forecast error (RFE), and mean absolute forecast error (MAFE), defined as
\begin{align*}
\mathrm{RMSFE} &= \sqrt{ \frac{1}{N_{\textrm{test}}} \sum_{\iota=1}^{N_{\text{test}}} \left( Y_{\iota}^* - \widehat{Y}_{\iota}^* \right)^2 },\\
\mathrm{RFE} &= \frac{1}{N_{\textrm{test}}} \sum_{\iota=1}^{N_{\mathrm{test}}} \left| \frac{Y_{\iota}^* - \widehat{Y}_{\iota}^*}{Y_{\iota}^*} \right|,\\
\mathrm{MAFE} &= \frac{1}{N_{\textrm{test}}} \sum_{\iota=1}^{N_{\mathrm{test}}} \left| Y_{\iota}^* - \widehat{Y}_{\iota}^* \right|,
\end{align*}
where the global prediction $\widehat{Y}_{\iota}^*$ is computed using~\eqref{eq:dist_test}.

For interval estimation, we compute the empirical coverage probability (ECP) and the interval score (IS). For the training data, the ECP is given by
\begin{equation*}
    \textrm{ECP}_{\textrm{train}} = \frac{1}{N_{\textrm{train}}} \sum_{i=1}^{N_{\textrm{train}}} \mathbbm{1} \left[\widehat{Y}_i - \gamma \widehat{\sigma} \le Y_i \le \widehat{Y}_i + \gamma \widehat{\sigma} \right],
\end{equation*}
where $\gamma$ and $\widehat{\sigma}$ are obtained from~\eqref{eq:dist_gamma} and~\eqref{eq:dist_sd}.  

For the testing data, the same $\gamma$ and $\widehat{\sigma}$ are employed, and the corresponding ECP is calculated as
\begin{equation*}
    \mathrm{ECP}_{\textrm{test}} = \frac{1}{N_{\mathrm{test}}} \sum_{\iota=1}^{N_{\mathrm{test}}} \mathbbm{1}\left[\widehat{Y}_{\iota}^* - \gamma \widehat{\sigma} \le Y_{\iota}^* \le \widehat{Y}_{\iota}^* + \gamma \widehat{\sigma}\right]. 
\end{equation*}

For a central $(1 - \alpha) \times 100\%$ prediction interval with lower bound $\widehat{Y}_{\iota}^* - \gamma \widehat{\sigma}$ and upper bound $\widehat{Y}_{\iota}^* + \gamma \widehat{\sigma}$, the interval score, as defined in Equation (43) of \citet{gneiting2007strictly}, is given by
\begin{align*}
    \mathrm{IS} = \frac{1}{N_{\mathrm{test}}} \sum_{\iota = 1}^{N_{\mathrm{test}}} \bigg\{ 2 \gamma \widehat{\sigma} 
    &+ \frac{2}{\alpha} (\widehat{Y}_{\iota}^* - \gamma \widehat{\sigma} - Y_{\iota}^*) \mathbbm{1} \left[ Y_{\iota}^* < \widehat{Y}_{\iota}^* - \gamma \widehat{\sigma} \right] \\ 
    &+ \frac{2}{\alpha} (Y_{\iota}^* - \widehat{Y}_{\iota}^* - \gamma \widehat{\sigma}) \mathbbm{1} \left[ Y_{\iota}^* > \widehat{Y}_{\iota}^* + \gamma \widehat{\sigma} \right] \bigg\}.
\end{align*}

\subsection{Results}
\label{sec:sim_results}

This section presents the results of the simulation studies. For each study, the full training sample consists of 2000 observations, which are partitioned into different numbers of blocks ranging from 1 to 40. The testing sample is randomly selected and fixed at 200, 400, or 800. All evaluation criteria are averaged over 200 Monte Carlo replications. 

\subsubsection{Scalar-on-Function Linear Model}\label{sec:sim_FLM}

Table~\ref{tbl:FLM_point_basis} reports the estimation and prediction performance of the FLM using the B-spline expansion method. As the number of blocks $K$ increases from 1 to 40, the sample size per block decreases from 2000 to 50. For the in-sample estimation, both RMSE and MAE exhibit a slight decrease as $K$ increases, suggesting a marginal improvement in the estimation accuracy of the response variable. In contrast, the estimation accuracy of the functional coefficient decreases slightly with larger $K$, as indicated by increases in the Frobenius norm, pointwise bias, and standard deviation. However, these increases are very small in magnitude, indicating that the loss of accuracy in estimating $\beta(t)$ remains limited even with substantial data partitioning. For the out-of-sample prediction, both RMSFE and MAFE remain relatively stable for each fixed testing sample size. The values exhibit only minor variation as $K$ increases, and this conclusion holds consistently across all testing sample sizes considered.
\begin{table}[!htb]
    \centering
    \tabcolsep 0.115in
    \caption{\small Estimation and prediction performance of the FLM using B-spline expansion method under varying numbers of blocks and testing sample sizes. All results are averaged over 200 Monte Carlo simulations.}
    \label{tbl:FLM_point_basis}
    \footnotesize
    \begin{tabular}{@{}ccccccccccc@{}}
        \toprule
        ~ & ~ & \multicolumn{6}{c}{Training data} & \multicolumn{3}{c}{Testing data} \\
        \cmidrule(lr){3-11}
        ~ & ~ & \multicolumn{3}{c}{Response variable} & \multicolumn{3}{c}{Functional coefficient} & \multicolumn{3}{c}{Response variable} \\
        \cmidrule(lr){3-5}\cmidrule(lr){6-8}\cmidrule(lr){9-11}
        $K$ & Time & $N_{\mathrm{train}}$ & RMSE & MAE & F.Norm & $\textrm{Bias}^2$ & ST.DEV & $N_{\mathrm{test}}$ & RMSFE & MAFE \\
        \midrule 
        1 & 2.8880s & 2000 & 0.9997 & 0.7978 & 0.5515 & 0.4775 & 0.2728 & 200 & 0.9962 & 0.7971 \\
        ~ & ~ & ~ & ~ & ~ & ~ & ~ & ~ & 400 & 0.9975 & 0.7956 \\
        ~ & ~ & ~ & ~ & ~ & ~ & ~ & ~ & 800 & 1.0029 & 0.8010 \\
        \midrule
        2 & 0.3728s & 1000 & 0.9982 & 0.7966 & 0.5516 & 0.4769 & 0.2740 & 200 & 0.9962 & 0.7971 \\    
        ~ & ~ & ~ & ~ & ~ & ~ & ~ & ~ & 400 & 0.9975 & 0.7956 \\
        ~ & ~ & ~ & ~ & ~ & ~ & ~ & ~ & 800 & 1.0029 & 0.8010 \\
        \midrule
        5 & 0.0332s & 400 & 0.9938 & 0.7932 & 0.5580 & 0.4823 & 0.2758 & 200 & 0.9962 & 0.7971 \\
        ~ & ~ & ~ & ~ & ~ & ~ & ~ & ~ & 400 & 0.9975 & 0.7957 \\
        ~ & ~ & ~ & ~ & ~ & ~ & ~ & ~ & 800 & 1.0030 & 0.8010 \\
        \midrule
        10 & 0.0092s & 200 & 0.9862 & 0.7870 & 0.5547 & 0.4769 & 0.2796 & 200 & 0.9961 & 0.7971 \\
        ~ & ~ & ~ & ~ & ~ & ~ & ~ & ~ & 400 & 0.9976 & 0.7957 \\
        ~ & ~ & ~ & ~ & ~ & ~ & ~ & ~ & 800 & 1.0030 & 0.8010 \\
        \midrule
        20 & 0.0051s & 100 & 0.9709 & 0.7748 & 0.5479 & 0.4683 & 0.2828 & 200 & 0.9963 & 0.7972 \\
        ~ & ~ & ~ & ~ & ~ & ~ & ~ & ~ & 400 & 0.9976 & 0.7957 \\
        ~ & ~ & ~ & ~ & ~ & ~ & ~ & ~ & 800 & 1.0030 & 0.8010 \\
        \midrule
        40 & 0.0041s & 50 & 0.9398 & 0.7498 & 0.5671 & 0.4780 & 0.2992 & 200 & 0.9964 & 0.7972 \\
        ~ & ~ & ~ & ~ & ~ & ~ & ~ & ~ & 400 & 0.9978 & 0.7958 \\
        ~ & ~ & ~ & ~ & ~ & ~ & ~ & ~ & 800 & 1.0032 & 0.8012 \\
        \bottomrule 
    \end{tabular} 
\end{table}

We omit RE and RFE for the FLM. In this simulation design, the responses $Y_i$ are centred around zero. Because both RE and RFE require division by $Y_i$, small values of $Y_i$ inflate the relative errors substantially, producing misleadingly large statistics.

Execution time is also reported as a measure of computational efficiency. We define it as the average time required to fit the model on a single block. Assuming $K$ machines are available to run the computations in parallel, this average serves as a reasonable proxy for computational efficiency. 
As $K$ increases and the block size decreases, the average execution time per block, as well as the total execution time\footnote{Here, ``total execution time'' refers to the scenario in which a large dataset is processed on a single machine, with the goal of reducing overall computation time. It can be approximated as the execution time per block multiplied by the number of blocks.} drops substantially, demonstrating the computational benefits of the distributed approach.

Table~\ref{tbl:FLM_ECP_basis} presents the ECP and IS for various values of $K$, for $\alpha = 0.05$ and $\alpha = 0.20$. When $K = 1$, the ECP for the training data is exactly equal to the nominal coverage probability $1 - \alpha$. As $K$ increases, the ECP decreases slightly below $1 - \alpha$. A similar pattern is observed for the testing data, although the magnitude of the decline in ECP is larger for the testing data than for the training data.
\begin{table}[!htb]
\centering
\tabcolsep 0.07in
\caption{\small Mean empirical coverage probability (ECP) and mean interval score (IS) of the prediction intervals for the FLM using B-spline expansion method under varying numbers of blocks. All results are averaged over 200 Monte Carlo simulations.}\label{tbl:FLM_ECP_basis}
\begin{tabular}{@{}lccccccccc@{}}
        \toprule
        & ~ & \multicolumn{2}{c}{Training Data} & \multicolumn{2}{c}{$N_{\mathrm{test}} = 200$} & \multicolumn{2}{c}{$N_{\mathrm{test}} = 400$} & \multicolumn{2}{c}{$N_{\mathrm{test}} = 800$} \\
        \cmidrule(lr){3-4}\cmidrule(lr){5-6}\cmidrule(lr){7-8}\cmidrule(lr){9-10}
        Criterion & $K$ & $\alpha = 0.05$ & $\alpha = 0.2$ & $\alpha = 0.05$ & $\alpha = 0.2$ & $\alpha = 0.05$ & $\alpha = 0.2$ & $\alpha = 0.05$ & $\alpha = 0.2$ \\
        \midrule 
        ECP & 1 & 95.00\% & 80.00\% & 94.97\% & 80.23\% & 94.87\% & 80.16\% & 94.91\% & 79.84\% \\
            & 2 & 94.95\% & 79.98\% & 94.94\% & 80.15\% & 94.82\% & 80.08\% & 94.86\% & 79.73\% \\
            & 5 & 94.87\% & 79.91\% & 94.75\% & 79.85\% & 94.66\% & 79.83\% & 94.67\% & 79.47\% \\
            & 10 & 94.76\% & 79.86\% & 94.53\% & 79.43\% & 94.42\% & 79.41\% & 94.40\% & 79.03\% \\
            & 20 & 94.53\% & 79.71\% & 93.90\% & 78.43\% & 93.77\% & 78.57\% & 93.72\% & 78.18\% \\
            & 40 & 94.06\% & 79.41\% & 92.53\% & 76.59\% & 92.37\% & 76.76\% & 92.27\% & 76.32\% \\
            \midrule 
        IS & 1 & 4.6707 & 3.5079 & 4.6678 & 3.4947 & 4.6896 & 3.5090 & 4.6911 & 3.5192 \\
        & 2 & 4.6646 & 3.5030 & 4.6683 & 3.4948 & 4.6893 & 3.5088 & 4.6915 & 3.5193 \\
        & 5 & 4.6428 & 3.4865 & 4.6690 & 3.4945 & 4.6907 & 3.5093 & 4.6926 & 3.5196 \\
        & 10 & 4.6086 & 3.4601 & 4.6702 & 3.4950 & 4.6946 & 3.5098 & 4.6970 & 3.5205 \\
        & 20 & 4.5433 & 3.4073 & 4.6862 & 3.4981 & 4.7140 & 3.5119 & 4.7164 & 3.5239 \\
        & 40 & 4.4087 & 3.2992 & 4.7536 & 3.5102 & 4.7868 & 3.5242 & 4.7910 & 3.5374 \\
        \bottomrule
    \end{tabular} 
\end{table}

For the IS, the behaviour differs between training and testing sets. When $K = 1$, the IS values for both sets are very similar. As $K$ increases, the IS for the training data decreases, indicating narrower prediction intervals and improved precision. However, the IS for the testing data increases steadily with $K$ across all testing sample sizes. Since RMSFE and MAFE remain essentially constant, the decline in ECP implies that the prediction intervals are becoming narrower. Consequently, more testing observations fall outside the intervals, directly affecting IS. 

Table~\ref{tbl:FLM_point_FPCA} reports the estimation and prediction performance of the FLM using the FPCA method. As with the B-spline approach, changes in all criteria remain small across different values of $K$. However, in comparison with the B-spline method, FPCA yields more accurate estimates of the functional coefficient, as reflected by the lower Frobenius norm, bias, and standard deviation. This improvement arises because the B-spline expansion provides a smoothed approximation to the curve, whereas FPCA captures more local variability by optimally representing the empirical covariance structure.
\begin{table}[!htb]
    \centering
    \caption{\small Estimation and prediction performance of the FLM using FPCA method under varying numbers of blocks and testing sample sizes. All results are averaged over 200 Monte Carlo simulations.}
    \label{tbl:FLM_point_FPCA}
    \tabcolsep 0.112in
    \footnotesize
    \begin{tabular}{@{}ccccccccccc@{}}
        \toprule
        ~ & ~ & \multicolumn{6}{c}{Training data} & \multicolumn{3}{c}{Testing data} \\
        \midrule
        ~ & ~ & \multicolumn{3}{c}{Response variable} & \multicolumn{3}{c}{Functional coefficient} & \multicolumn{3}{c}{Response variable} \\
        \cmidrule(lr){3-5}\cmidrule(lr){6-8}\cmidrule(lr){9-11}
        $K$ & Time & $N_{\mathrm{train}}$ & RMSE & MAE & F.Norm & $\textrm{Bias}^2$ & ST.DEV & $N_{\mathrm{test}}$ & RMSFE & MAFE \\
        \midrule 
        1 & 0.7949s & 2000 & 0.9997 & 0.7978 & 0.1550 & 0.1512 & 0.0619 & 200 & 0.9962 & 0.7971 \\
        ~ & ~ & ~ & ~ & ~ & ~ & ~ & ~ & 400 & 0.9975 & 0.7956 \\
        ~ & ~ & ~ & ~ & ~ & ~ & ~ & ~ & 800 & 1.0029 & 0.8010 \\
        \midrule
        2 & 0.3076s & 1000 & 0.9982 & 0.7966 & 0.1550 & 0.1512 & 0.0620 & 200 & 0.9962 & 0.7971 \\
        ~ & ~ & ~ & ~ & ~ & ~ & ~ & ~ & 400 & 0.9975 & 0.7956 \\
        ~ & ~ & ~ & ~ & ~ & ~ & ~ & ~ & 800 & 1.0029 & 0.8010 \\
        \midrule
        5 & 0.1533s & 400 & 0.9938 & 0.7932 & 0.1551 & 0.1512 & 0.0628 & 200 & 0.9962 & 0.7971 \\
        ~ & ~ & ~ & ~ & ~ & ~ & ~ & ~ & 400 & 0.9975 & 0.7956 \\
        ~ & ~ & ~ & ~ & ~ & ~ & ~ & ~ & 800 & 1.0030 & 0.8010 \\
        \midrule
        10 & 0.0896s & 200 & 0.9862 & 0.7870 & 0.1551 & 0.1512 & 0.0631 & 200 & 0.9961 & 0.7971 \\
        ~ & ~ & ~ & ~ & ~ & ~ & ~ & ~ & 400 & 0.9976 & 0.7957 \\
        ~ & ~ & ~ & ~ & ~ & ~ & ~ & ~ & 800 & 1.0030 & 0.8010 \\
        \midrule
        20 & 0.0561s & 100 & 0.9709 & 0.7748 & 0.1552 & 0.1512 & 0.0639 & 200 & 0.9963 & 0.7972 \\
        ~ & ~ & ~ & ~ & ~ & ~ & ~ & ~ & 400 & 0.9976 & 0.7957 \\
        ~ & ~ & ~ & ~ & ~ & ~ & ~ & ~ & 800 & 1.0030 & 0.8010 \\
        \midrule
        40 & 0.0299s & 50 & 0.9398 & 0.7498 & 0.1556 & 0.1512 & 0.0669 & 200 & 0.9964 & 0.7972 \\
        ~ & ~ & ~ & ~ & ~ & ~ & ~ & ~ & 400 & 0.9978 & 0.7958 \\
        ~ & ~ & ~ & ~ & ~ & ~ & ~ & ~ & 800 & 1.0032 & 0.8012 \\
        \bottomrule 
    \end{tabular} 
\end{table}

A more substantial difference between the two methods appears in execution time. With $K = 1$ (block size 2000), FPCA is considerably faster (approximately 0.8 seconds) than the B-spline method (approximately 2.9 seconds). As $K$ increases, the execution time for the B-spline method decreases rapidly, with approximately a ten-fold reduction when $K$ doubles, whereas the execution time for FPCA decreases only moderately, roughly halving as $K$ doubles. This difference reflects the higher computational overhead of spline fitting relative to computing a small number of principal components in our simulation studies. 

Table~\ref{tbl:FLM_ECP_FPCA} reports the mean ECP and IS for the FLM using FPCA. As with the B-spline method, increases in $K$ lead to modest changes in both metrics. The ECP decreases slightly, while IS increases slightly, consistent with narrower prediction intervals and reduced empirical coverage.
\begin{table}[!htb]
\centering
\tabcolsep 0.067in
\caption{\small Mean empirical coverage probability (ECP) and mean interval score (IS) of the prediction intervals for the FLM using FPCA method under varying numbers of blocks. All results are averaged over 200 Monte Carlo simulations.}
\label{tbl:FLM_ECP_FPCA}
\begin{tabular}{@{}lccccccccc@{}}
\toprule
& ~ & \multicolumn{2}{c}{Training Data} & \multicolumn{2}{c}{$N_{\mathrm{test}} = 200$} & \multicolumn{2}{c}{$N_{\mathrm{test}} = 400$} & \multicolumn{2}{c}{$N_{\mathrm{test}} = 800$} \\
\cmidrule(lr){3-4}\cmidrule(lr){5-6}\cmidrule(lr){7-8}\cmidrule(lr){9-10}
Criterion & $K$ & $\alpha = 0.05$ & $\alpha = 0.2$ & $\alpha = 0.05$ & $\alpha = 0.2$ & $\alpha = 0.05$ & $\alpha = 0.2$ & $\alpha = 0.05$ & $\alpha = 0.2$ \\
\midrule 
ECP & 1 & 95.00\% & 80.00\% & 94.97\% & 80.23\% & 94.87\% & 80.16\% & 94.91\% & 79.84\% \\
& 2 & 94.95\% & 79.98\% & 94.94\% & 80.15\% & 94.82\% & 80.08\% & 94.86\% & 79.73\% \\
& 5 & 94.87\% & 79.91\% & 94.75\% & 79.85\% & 94.66\% & 79.83\% & 94.67\% & 79.47\% \\
& 10 & 94.76\% & 79.86\% & 94.53\% & 79.43\% & 94.42\% & 79.41\% & 94.40\% & 79.03\% \\
& 20 & 94.53\% & 79.71\% & 93.90\% & 78.43\% & 93.77\% & 78.57\% & 93.72\% & 78.18\% \\
& 40 & 94.06\% & 79.41\% & 92.53\% & 76.60\% & 92.37\% & 76.76\% & 92.27\% & 76.32\% \\
\midrule
IS & 1 & 4.6707 & 3.5079 & 4.6678 & 3.4947 & 4.6896 & 3.5090 & 4.6911 & 3.5192 \\
& 2 & 4.6646 & 3.5030 & 4.6683 & 3.4948 & 4.6893 & 3.5088 & 4.6915 & 3.5193 \\
& 5 & 4.6428 & 3.4865 & 4.6690 & 3.4945 & 4.6907 & 3.5093 & 4.6926 & 3.5196 \\
& 10 & 4.6086 & 3.4601 & 4.6702 & 3.4950 & 4.6946 & 3.5098 & 4.6970 & 3.5205 \\
& 20 & 4.5433 & 3.4073 & 4.6862 & 3.4981 & 4.7140 & 3.5119 & 4.7164 & 3.5239 \\
& 40 & 4.4087 & 3.2992 & 4.7536 & 3.5102 & 4.7868 & 3.5242 & 4.7910 & 3.5374 \\
\bottomrule
\end{tabular} 
\end{table}

Overall, for the FLM, increasing the number of blocks yields modest declines in both in-sample estimation accuracy and out-of-sample predictive performance, along with slightly narrower prediction intervals. However, these losses are very small, whereas the reduction in computation time is substantial. Thus, distributed estimation provides considerable gains in efficiency with only minimal degradation in statistical performance.

\FloatBarrier

\subsubsection{Scalar-on-function Non-Parametric Model}
\label{sec:sim_FNPM}

Table~\ref{tbl:FNPM_point} summarises the estimation and prediction performance of the FNPM. For the in-sample estimation on the training data, all three evaluation criteria (RMSE, RE, and MAE) show a slight improvement as the number of blocks increases. As $K$ increases from 1 to 40, the RMSE decreases from 1.3971 to 1.2776, the RE declines from 8.34\% to 7.53\%, and the MAE decreases from 1.1147 to 1.0073. These results indicate that partitioning the data into smaller blocks leads to marginal gains in fitting accuracy for the training sample.
\begin{table}[!htb]
\centering
\tabcolsep 0.135in
    \caption{\small Estimation and prediction performance of the FNPM under varying numbers of blocks and testing sample sizes. All results are averaged over 200 Monte Carlo simulations.}
    \label{tbl:FNPM_point}
    \begin{tabular}{@{}cccccccccc@{}}
        \toprule
        ~ & ~ & \multicolumn{8}{c}{Response variable} \\
        \cmidrule(lr){3-10}
        ~ & ~ & \multicolumn{4}{c}{Training data} & \multicolumn{4}{c}{Testing data} \\
        \cmidrule(lr){3-6}\cmidrule(lr){7-10}
        $K$ & Time & $N_{\mathrm{train}}$ & RMSE & RE & MAE & $N_{\mathrm{test}}$ & RMSFE & RFE & MAFE \\
        \midrule 
        1 & 29.37s & 2000 & 1.3971 & 8.34\% & 1.1147 & 200 & 1.4182 & 8.49\% & 1.1341 \\
        ~ & ~ & ~ & ~ & ~ & ~ & 400 & 1.4365 & 8.60\% & 1.1478 \\
        ~ & ~ & ~ & ~ & ~ & ~ & 800 & 1.4335 & 8.56\% & 1.1429 \\
        \midrule
        2 & 7.61s & 1000 & 1.3757 & 8.21\% & 1.0977 & 200 & 1.4190 & 8.49\% & 1.1348 \\
        ~ & ~ & ~ & ~ & ~ & ~ & 400 & 1.4373 & 8.60\% & 1.1484 \\
        ~ & ~ & ~ & ~ & ~ & ~ & 800 & 1.4343 & 8.56\% & 1.1434 \\
        \midrule
        5 & 1.41s & 400 & 1.3494 & 8.05\% & 1.0762 & 200 & 1.4249 & 8.53\% & 1.1395 \\
        ~ & ~ & ~ & ~ & ~ & ~ & 400 & 1.4428 & 8.64\% & 1.1528 \\
        ~ & ~ & ~ & ~ & ~ & ~ & 800 & 1.4400 & 8.60\% & 1.1480 \\
        \midrule
        10 & 0.46s & 200 & 1.3324 & 7.94\% & 1.0616 & 200 & 1.4347 & 8.59\% & 1.1477 \\
        ~ & ~ & ~ & ~ & ~ & ~ & 400 & 1.4524 & 8.70\% & 1.1603 \\
        ~ & ~ & ~ & ~ & ~ & ~ & 800 & 1.4495 & 8.66\% & 1.1556 \\
        \midrule
        20 & 0.16s & 100 & 1.3170 & 7.82\% & 1.0456 & 200 & 1.4494 & 8.69\% & 1.1600 \\
        ~ & ~ & ~ & ~ & ~ & ~ & 400 & 1.4666 & 8.78\% & 1.1717 \\
        ~ & ~ & ~ & ~ & ~ & ~ & 800 & 1.4641 & 8.75\% & 1.1675 \\
        \midrule
        40 & 0.07s & 50 & 1.2776 & 7.53\% & 1.0073 & 200 & 1.4612 & 8.76\% & 1.1698 \\
        ~ & ~ & ~ & ~ & ~ & ~ & 400 & 1.4783 & 8.85\% & 1.1811 \\
        ~ & ~ & ~ & ~ & ~ & ~ & 800 & 1.4758 & 8.82\% & 1.1769 \\
        \bottomrule
    \end{tabular} 
\end{table}

\vspace{.2in}

For the out-of-sample prediction, the differences across block sizes are much smaller. Prediction accuracy declines only slightly as $K$ increases. For example, with a testing sample size of 800, increasing $K$ from 1 to 40 results in only minor increases in the prediction errors: RMSFE rises from 1.4335 to 1.4758, RFE from 8.56\% to 8.82\%, and MAFE from 1.1429 to 1.1769. Overall, the FNPM becomes slightly more accurate in fitting the training data after partitioning, but the predictive errors increase marginally.

The most substantial benefit of distributed estimation for the FNPM appears in execution time. As $K$ doubles, the execution time decreases by a factor of four. This behaviour is fully consistent with the theoretical computational complexity of the kernel estimator in~\eqref{eq:est_FNPM}. To estimate $\widehat{m}(\mathcal{X}_j)$ for a single functional predictor $\mathcal{X}_j$, the computation proceeds in three steps:
\begin{itemize}
    \item[1)] Compute the semi-metric $d(\mathcal{X}_j, \mathcal{X}_i)$ in \eqref{eq:dist} for all $i = 1, \dots, n$, which requires $\mathcal{O}(Mn)$ operations.
    \item[2)] Evaluate kernel values $K(h^{-1} d(\mathcal{X}_j, \mathcal{X}_i))$, which requires $\mathcal{O}(n)$ operations. 
    \item[3)] Compute the weighted average to obtain $\widehat{m}(\mathcal{X}_j)$, which also requires $\mathcal{O}(n)$ operations.
\end{itemize}
Thus, the total complexity for estimating one function is $\mathcal{O}(Mn)$, and for all $n$ observations in a block, it becomes $\mathcal{O}(Mn^2)$. If $K$ doubles, the block size $n$ is halved, and hence the total computational burden falls by a factor of four. The execution time observed in Table~\ref{tbl:FNPM_point} aligns closely with this theoretical prediction.

Table~\ref{tbl:FNPM_ECP} reports the mean ECP and IS for different numbers of blocks. As $K$ increases, the ECP decreases for both the training and testing data, with the decline being larger for the testing data. For the training data, the IS decreases slightly as $K$ increases, indicating narrower intervals and improved in-sample precision.
However, for the testing data, the IS increases with $K$, mirroring the behaviour observed for the FLM. Because the prediction errors remain relatively stable across values of $K$, the decline in ECP implies narrower intervals, and the higher IS simply reflects the larger number of testing observations that fall outside these intervals.
\begin{table}[!htb]
    \centering
    \tabcolsep 0.068in
    \caption{\small Mean empirical coverage probability (ECP) and mean interval score (IS) of the prediction intervals for the FNPM under varying numbers of blocks. All results are averaged over 200 Monte Carlo simulations.}
    \label{tbl:FNPM_ECP}
    \begin{tabular}{@{}lccccccccc@{}}
        \toprule
        & ~ & \multicolumn{2}{c}{Training Data} & \multicolumn{2}{c}{$N_{\mathrm{test}} = 200$} & \multicolumn{2}{c}{$N_{\mathrm{test}} = 400$} & \multicolumn{2}{c}{$N_{\mathrm{test}} = 800$} \\
        \cmidrule(lr){3-4}\cmidrule(lr){5-6}\cmidrule(lr){7-8}\cmidrule(lr){9-10}
        Criterion & $K$ & $\alpha = 0.05$ & $\alpha = 0.2$ & $\alpha = 0.05$ & $\alpha = 0.2$ & $\alpha = 0.05$ & $\alpha = 0.2$ & $\alpha = 0.05$ & $\alpha = 0.2$ \\
        \midrule 
        ECP & 1 & 95.00\% & 80.00\% & 94.57\% & 79.22\% & 94.34\% & 78.70\% & 94.30\% & 78.85\% \\
        & 2 & 94.95\% & 79.97\% & 94.12\% & 78.43\% & 93.87\% & 77.91\% & 93.85\% & 78.10\% \\
        & 5 & 94.87\% & 79.92\% & 93.49\% & 77.20\% & 93.15\% & 76.78\% & 93.17\% & 76.94\% \\
        & 10 & 94.71\% & 79.80\% & 92.79\% & 76.24\% & 92.40\% & 75.61\% & 92.45\% & 75.92\% \\
        & 20 & 94.32\% & 79.64\% & 91.83\% & 74.83\% & 91.42\% & 74.25\% & 91.51\% & 74.51\% \\
        & 40 & 93.63\% & 79.29\% & 89.83\% & 72.40\% & 89.44\% & 71.81\% & 89.56\% & 72.11\% \\
        \midrule
        IS & 1 & 6.5295 & 4.9036 & 6.6289 & 4.9832 & 6.7217 & 5.0442 & 6.7396 & 5.0402 \\
        & 2 & 6.4300 & 4.8282 & 6.6488 & 4.9898 & 6.7420 & 5.0515 & 6.7646 & 5.0475 \\
        & 5 & 6.3138 & 4.7379 & 6.7090 & 5.0203 & 6.8110 & 5.0832 & 6.8341 & 5.0778 \\
        & 10 & 6.2558 & 4.6818 & 6.8050 & 5.0676 & 6.9166 & 5.1325 & 6.9354 & 5.1261 \\
        & 20 & 6.2412 & 4.6403 & 6.9604 & 5.1411 & 7.0821 & 5.2077 & 7.1008 & 5.2012 \\
        & 40 & 6.1514 & 4.5262 & 7.2625 & 5.2339 & 7.4034 & 5.3068 & 7.4193 & 5.2978 \\
        \bottomrule
    \end{tabular} 
\end{table}

\subsubsection{Scalar-on-Function Partial Linear Model}
\label{sec:sim_FPLM}

Table~\ref{tbl:FPLM_point} summarises the estimation and prediction performance of the FPLM. Similar to the FLM and FNPM, increasing the number of blocks $K$ has only a small effect on the out-of-sample prediction errors. For all testing sample sizes, the RMSFE, RFE, and MAFE increase only marginally as $K$ increases. In contrast, the behaviour of the in-sample estimation errors is markedly different from the previous two models. As $K$ increases from 1 to 40, all three criteria (RMSE, RE, and MAE) decrease sharply, falling to roughly half of their original values. For example, the RMSE declines from 1.3842 to 0.7094, the RE from 8.27\% to 3.87\%, and the MAE from 1.1044 to 0.5183. Such a substantial improvement in in-sample fit, coupled with only mild changes in out-of-sample accuracy, strongly suggests the presence of overfitting when the training sample size is small. 
\begin{table}[!htb]
    \centering
    \tabcolsep 0.13in
    \renewcommand{\arraystretch}{0.82}
    \caption{\small Estimation and prediction performance of the FPLM under varying numbers of blocks and testing sample sizes. All results are averaged over 200 Monte Carlo simulations.}
    \label{tbl:FPLM_point}
    \begin{tabular}{@{}cccccccccc@{}}
        \toprule
        ~ & ~ & \multicolumn{8}{c}{Response variable} \\
        \cmidrule{3-10}
        ~ & ~ & \multicolumn{4}{c}{Training Data} & \multicolumn{4}{c}{Testing Data} \\
        \cmidrule(lr){3-6}\cmidrule(lr){7-10}
        $K$ & Time & $N_{\mathrm{train}}$ & RMSE & RE & MAE & $N_{\mathrm{test}}$ & RMSFE & RFE & MAFE \\
        \midrule 
        1 & 185.71s & 2000 & 1.3842 & 8.27\% & 1.1044 & 200 & 1.4362 & 8.57\% & 1.1449 \\
        ~ & ~ & ~ & ~ & ~ & ~ & 400 & 1.4302 & 8.56\% & 1.1431 \\
        ~ & ~ & ~ & ~ & ~ & ~ & 800 & 1.4307 & 8.55\% & 1.1420 \\
        \midrule
        2 & 24.95s & 1000 & 1.3479 & 8.05\% & 1.0755 & 200 & 1.4363 & 8.57\% & 1.1450 \\
        ~ & ~ & ~ & ~ & ~ & ~ & 400 & 1.4305 & 8.56\% & 1.1435 \\
        ~ & ~ & ~ & ~ & ~ & ~ & 800 & 1.4307 & 8.55\% & 1.1420 \\
        \midrule
        5 & 2.65s & 400 & 1.2490 & 7.45\% & 0.9948 & 200 & 1.4368 & 8.58\% & 1.1453 \\
        ~ & ~ & ~ & ~ & ~ & ~ & 400 & 1.4308 & 8.56\% & 1.1436 \\
        ~ & ~ & ~ & ~ & ~ & ~ & 800 & 1.4310 & 8.55\% & 1.1423 \\
        \midrule
        10 & 0.61s & 200 & 1.1149 & 6.59\% & 0.8817 & 200 & 1.4375 & 8.58\% & 1.1457 \\
        ~ & ~ & ~ & ~ & ~ & ~ & 400 & 1.4315 & 8.57\% & 1.1442 \\
        ~ & ~ & ~ & ~ & ~ & ~ & 800 & 1.4318 & 8.56\% & 1.1429 \\
        \midrule
        20 & 0.19s & 100 & 0.9260 & 5.35\% & 0.7157 & 200 & 1.4389 & 8.59\% & 1.1466 \\
        ~ & ~ & ~ & ~ & ~ & ~ & 400 & 1.4328 & 8.57\% & 1.1453 \\
        ~ & ~ & ~ & ~ & ~ & ~ & 800 & 1.4332 & 8.57\% & 1.1439 \\
        \midrule
        40 & 0.07s & 50 & 0.7094 & 3.87\% & 0.5183 & 200 & 1.4430 & 8.62\% & 1.1503 \\
        ~ & ~ & ~ & ~ & ~ & ~ & 400 & 1.4367 & 8.60\% & 1.1486 \\
        ~ & ~ & ~ & ~ & ~ & ~ & 800 & 1.4370 & 8.59\% & 1.1469 \\
        \bottomrule
    \end{tabular} 
\end{table}

As with the FLM and FNPM, distributed learning offers considerable computational advantages. The execution time decreases dramatically, from nearly 200 seconds with a single block to just 0.07 seconds with 40 blocks. This empirical reduction is well aligned with the theoretical computational complexity of the FPLM estimators in~\eqref{eq:beta_est_partial} and~\eqref{eq:est_partial}. The estimation procedure involves the following steps:
\begin{itemize}
    \item[1)] Evaluate the semi-metric $d(\mathcal{X}_j, \mathcal{X}_i)$ in \eqref{eq:dist} for all $i = 1,\dots,n$, with complexity $\mathcal{O}(Mn)$.
    \item[2)] Compute $w_h(\mathcal{X}_j, \mathcal{X}_i)$ for all $i$, with complexity $\mathcal{O}(n)$, giving a combined complexity of $\mathcal{O}(Mn)$ for Steps~1) and~2).
    \item[3)] Build the $n \times n$ matrix $\boldsymbol{W}_h$. Since $\boldsymbol{W}_h$ is symmetric, the number of unique distances is $\frac{n(n+1)}{2}$, and each requires $\mathcal{O}(M)$ operations. This results in a total complexity of~$\mathcal{O}(Mn^3)$.
    \item[4)] Compute $\widetilde{\boldsymbol{Z}} = (\boldsymbol{I} - \boldsymbol{W}_h) \boldsymbol{Y}$ and $\widetilde{\boldsymbol{Y}}_h = (\boldsymbol{I} - \boldsymbol{W}_h) \boldsymbol{Y}$, each with complexity $\mathcal{O}(n^2)$.
    \item[5)] Estimate $\widehat{\beta}_{NF}$. Since $\boldsymbol{Z}$ is one-dimensional, this step has a complexity of $\mathcal{O}(n)$.
    \item[6)] Evaluate $\widehat{m}(\mathcal{X})$ for all functions, which has complexity $\mathcal{O}(Mn^2)$ as in the FNPM.
\end{itemize}
The dominant computational burden arises from Steps 3), of order $\mathcal{O}(Mn^2)$. Consequently, when $K$ doubles, the block size $n$ halves, and the overall execution time is reduced by a factor of approximately eight. The empirical results in Table~\ref{tbl:FPLM_point} closely follow this pattern.

Table~\ref{tbl:FPLM_ECP} reports the mean ECP and IS for different values of $K$. As $K$ increases, the training ECP decreases slightly. However, because the in-sample estimation errors decline sharply, the IS for the training data decreases substantially, indicating much narrower prediction intervals. The interval width, given by $\gamma \widehat{\sigma}$, decreases correspondingly. Given that the out-of-sample prediction errors remain similar across all block sizes, narrower intervals lead to a much lower ECP for the testing data, as more test observations fall outside the interval. This decrease in ECP contributes directly to a decrease in the IS for the testing data as $K$ increases.
\begin{table}[!htb]
    \centering
    \tabcolsep 0.068in
    \caption{\small Mean empirical coverage probability (ECP) and mean interval score (IS) of the prediction intervals for the FPLM under varying numbers of blocks. All results are averaged over 200 Monte Carlo simulations.}
    \label{tbl:FPLM_ECP}
    \begin{tabular}{@{}lccccccccc@{}}
        \toprule
        ~ & ~ & \multicolumn{2}{c}{Training Data} & \multicolumn{2}{c}{$N_{\mathrm{test}} = 200$} & \multicolumn{2}{c}{$N_{\mathrm{test}} = 400$} & \multicolumn{2}{c}{$N_{\mathrm{test}} = 800$} \\
        \cmidrule(lr){3-4}\cmidrule(lr){5-6}\cmidrule(lr){7-8}\cmidrule(lr){9-10}
        Criterion & $K$ & $\alpha = 0.05$ & $\alpha = 0.2$ & $\alpha = 0.05$ & $\alpha = 0.2$ & $\alpha = 0.05$ & $\alpha = 0.2$ & $\alpha = 0.05$ & $\alpha = 0.2$ \\
        \midrule 
        ECP & 1 & 95.00\% & 80.00\% & 93.97\% & 78.33\% & 94.10\% & 78.40\% & 94.16\% & 78.43\% \\
        & 2 & 94.94\% & 79.98\% & 93.26\% & 77.03\% & 93.40\% & 77.15\% & 93.45\% & 77.18\% \\
        & 5 & 94.87\% & 79.92\% & 91.06\% & 73.28\% & 91.04\% & 73.35\% & 91.15\% & 73.45\% \\
        & 10 & 94.71\% & 79.84\% & 86.91\% & 67.49\% & 87.15\% & 67.61\% & 87.15\% & 67.70\% \\
        & 20 & 94.47\% & 79.70\% & 79.52\% & 57.93\% & 79.64\% & 58.05\% & 79.73\% & 58.02\% \\
        & 40 & 94.03\% & 79.61\% & 67.25\% & 44.96\% & 67.42\% & 44.77\% & 67.52\% & 44.93\% \\
        \midrule
        IS & 1 & 6.4686 & 4.8576 & 6.7802 & 5.0580 & 6.6952 & 5.0289 & 6.7099 & 5.0284 \\
        & 2 & 6.3033 & 4.7306 & 6.8213 & 5.0690 & 6.7316 & 5.0406 & 6.7438 & 5.0391 \\
        & 5 & 5.8656 & 4.3890 & 7.0411 & 5.1357 & 6.9527 & 5.1038 & 6.9530 & 5.1017 \\
        & 10 & 5.3110 & 3.9386 & 7.6674 & 5.3106 & 7.5641 & 5.2763 & 7.5631 & 5.2729 \\
        & 20 & 4.5729 & 3.3209 & 9.3085 & 5.7605 & 9.1775 & 5.7211 & 9.1830 & 5.7188 \\
        & 40 & 3.7557 & 2.6144 & 12.8454 & 6.6344 & 12.7013 & 6.5964 & 12.696 & 6.5911 \\
        \bottomrule
    \end{tabular} 
\end{table}

\section{\textit{Tecator} Data Analysis}
\label{sec:real_data}

\begin{figure}[!htb]
\centering
\includegraphics[width=11cm]{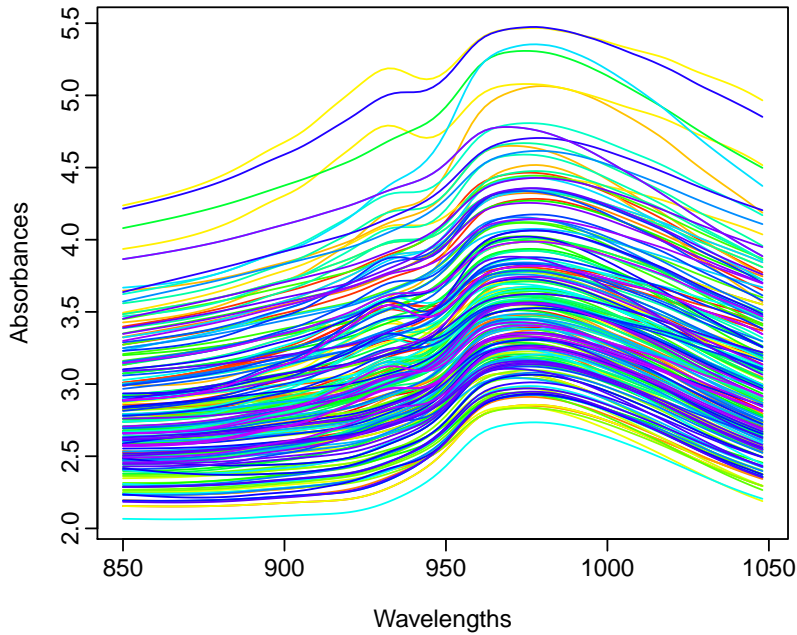}
\caption{A graphical display of spectrometric curves.}
\label{fig:spec}
\end{figure}

We consider a food quality control application, studied by \cite{ferraty2006nonparametric} and \cite{aneiros2006semi}. The data set was obtained from \url{https://lib.stat.cmu.edu/datasets/tecator}.  
For each unit~$i$ (among 215 pieces of finely chopped meat), we observe a spectrometric curve, denoted by $\X_i$, which corresponds to the absorbance measured on a grid of 100 wavelengths (i.e., $\X_i = (\X_i(t_1),\dots,\X_i(t_{100}))$). For each unit $i$, we also observe its fat/protein/moisture content $Y_i\in \mathbb{R}$ obtained by analytical chemical processing. The data set contains the pairs $(\X_i, Y_i)_{i=1}^{215}$. Given a new spectrometric curve $\X$, our task is to predict the corresponding fat/protein/moisture content. As pointed out by \cite{ferraty2006nonparametric}, the motivation is that obtaining a spectrometric curve is less time consuming than the analytic chemistry needed for determining the fat/protein/moisture content. A graphical display of the spectrometric curves is shown in Figure~\ref{fig:spec}.

We study the relationship between the spectrometric curves and its corresponding fat, protein, or moisture content, respectively. We use the nonparametric functional Nadaraya-Watson estimator. To assess the out-of-sample prediction accuracy of the nonparametric functional estimator, we split the original samples into three subsamples. The first one is called training sample, which contains the first 129 units $\{(\X_i, y_i)_{i=1,\dots,129}\}$. The second one is called validation sample, which contains the non-overlapping 43 units $\{(\X_i, y_i)_{i=130,\dots,172}$. The training sample allows us to build the functional NW estimator. To measure the prediction accuracy, we evaluate the functional NW estimator using the validation sample, from which we predict response $(y_{130},\dots,y_{172})$. We compute the residual $\epsilon_{\omega} = y_{\omega}-\widehat{y}_{\omega}, \omega = 1,\dots,43$.

In this section, we apply the distributed estimation procedures for the FLM, FNPM, and FPLM to the \textit{tecator} dataset. The full dataset contains 215 observations. We randomly select 150 observations as the training set and use the remaining 65 observations as the testing set. Each model is fitted on the training data and evaluated on the testing data using the criteria described in Section~\ref{sec:sim_acc}. To obtain stable results, this sampling procedure is repeated 200 times, and all reported statistics are averaged over these 200 samples.

Table~\ref{tbl:real_point} reports the estimation and prediction performance of all models. For the FPLM, although two non-functional covariates could be incorporated, we consider only a single non-functional covariate in this analysis. The number of blocks is increased up to a maximum of five, reducing the sample size per block from 150 to 30. Across all three models, increasing the number of blocks leads to a substantial reduction in execution time, while prediction errors remain relatively stable.

\begin{center}
\tabcolsep 0.056in
\renewcommand{\arraystretch}{0.85}
\begin{longtable}{@{}lcccccccccc@{}}
\caption{\small Estimation and prediction performance for all models under varying numbers of blocks. All results are averaged over 200 Monte Carlo replications.}\label{tbl:real_point} \\
        \toprule
        ~ & ~ & ~ & \multicolumn{4}{c}{Training Data} & \multicolumn{4}{c}{Testing Data} \\
        \midrule
        Model & $K$ & Time & $N_{\mathrm{train}}$ & RMSE & RE & MAE & $N_{\mathrm{test}}$ & RMSFE & RFE & MAFE \\
        \midrule 
\endfirsthead
        \toprule 
        ~ & ~ & ~ & \multicolumn{4}{c}{Training Data} & \multicolumn{4}{c}{Testing Data} \\
        \midrule
        Model & $K$ & Time & $N_{\mathrm{train}}$ & RMSE & RE & MAE & $N_{\mathrm{test}}$ & RMSFE & RFE & MAFE \\
        \midrule 
\endhead        
        \multicolumn{8}{l}{\underline{\hspace{-.075in}{\textit{Response variable: fat, non-functional covariate: moisture}}}} \\                
        FLM (B-spline) & 1 & 0.0074s & 150 & 4.7541 & 50.13\% & 3.8388 & 65 & 5.0195 & 52.26\% & 4.0497 \\
        ~& 2 & 0.0045s & 75 & 4.6323 & 48.40\% & 3.7307 & 65 & 5.0337 & 52.16\% & 4.0605 \\
        ~ & 3 & 0.0040s & 50 & 4.4984 & 46.62\% & 3.6181 & 65 & 5.0519 & 52.00\% & 4.0739 \\
        ~ & 5 & 0.0038s & 30 & 4.2573 & 43.51\% & 3.4140 & 65 & 5.0983 & 52.03\% & 4.1015 \\
        \midrule
        FLM (FPCA) & 1 & 0.1155s & 150 & 3.2881 & 28.44\% & 2.5964 & 65 & 3.5582 & 30.25\% & 2.7960 \\
        ~ & 2 & 0.0393s & 75 & 3.1820 & 27.43\% & 2.5194 & 65 & 3.5533 & 30.06\% & 2.7771 \\
        ~ & 3 & 0.0268s & 50 & 3.0709 & 26.38\% & 2.4324 & 65 & 3.5571 & 29.97\% & 2.7671 \\
        ~ & 5 & 0.0195s & 30 & 2.8676 & 24.62\% & 2.2802 & 65 & 3.5701 & 29.87\% & 2.7551 \\
        \midrule
        FNPM & 1 & 0.2918s & 150 & 5.5555 & 44.92\% & 4.3421 & 65 & 8.9223 & 65.56\% & 6.9614 \\
        ~ & 2 & 0.0880s & 75 & 5.2692 & 40.55\% & 3.9620 & 65 & 9.2352 & 70.77\% & 7.3446 \\
        ~ & 3 & 0.0536s & 50 & 5.9602 & 43.88\% & 4.3509 & 65 & 9.7325 & 76.19\% & 7.8627 \\
        ~ & 5 & 0.0363s & 30 & 7.0819 & 51.06\% & 5.1269 & 65 & 10.3648 & 82.76\% & 8.4778 \\
        \midrule
        FPLM & 1 & 0.3061s & 150 & 1.1140 & 7.97\% & 0.7505 & 65 & 1.8035 & 12.89\% & 1.2856 \\
        ~ & 2 & 0.0978s & 75 & 0.9102 & 6.28\% & 0.5927 & 65 & 1.8652 & 13.67\% & 1.3570 \\
        ~ & 3 & 0.0638s & 50 & 0.7613 & 5.08\% & 0.4858 & 65 & 1.9107 & 14.06\% & 1.4010 \\
        ~ & 5 & 0.0378s & 30 & 0.5904 & 3.72\% & 0.3590 & 65 & 2.0126 & 14.56\% & 1.4623 \\
        \midrule 
\multicolumn{8}{l}{\underline{\hspace{-.075in}{\textit{Response variable: moisture, non-functional covariate: protein}}}} \\
        FLM (B-spline) & 1 & 0.0102s & 150 & 3.6828 & 4.81\% & 2.9670 & 65 & 3.8718 & 5.06\% & 3.1136 \\
        ~& 2 & 0.0047s & 75 & 3.5916 & 4.68\% & 2.8848 & 65 & 3.8838 & 5.08\% & 3.1191 \\
        ~ & 3 & 0.0040s & 50 & 3.4917 & 4.54\% & 2.7999 & 65 & 3.9008 & 5.10\% & 3.1269 \\
        ~ & 5 & 0.0037s & 30 & 3.3063 & 4.29\% & 2.6432 & 65 & 3.9371 & 5.14\% & 3.1438 \\
        \midrule
        FLM (FPCA) & 1 & 0.1192s & 150 & 2.7573 & 3.79\% & 2.2249 & 65 & 2.9339 & 4.00\% & 2.3368 \\
        ~& 2 & 0.0421s & 75 & 2.6815 & 3.67\% & 2.1584 & 65 & 2.9413 & 3.99\% & 2.3247 \\
        ~ & 3 & 0.0270s & 50 & 2.6012 & 3.56\% & 2.0929 & 65 & 2.9551 & 3.98\% & 2.3170 \\
        ~ & 5 & 0.0187s & 30 & 2.4436 & 3.32\% & 1.9620 & 65 & 2.9783 & 3.97\% & 2.3016 \\
        \midrule
        FNPM & 1 & 0.2644s & 150 & 4.2591 & 5.70\% & 3.3938 & 65 & 6.7548 & 9.27\% & 5.3897 \\
        ~& 2 & 0.0871s & 75 & 3.9595 & 5.08\% & 3.0270 & 65 & 6.9662 & 9.78\% & 5.6628 \\
        ~ & 3 & 0.0576s & 50 & 4.4698 & 5.61\% & 3.3168 & 65 & 7.3446 & 10.53\% & 6.0678 \\
        ~ & 5 & 0.0365s & 30 & 5.3939 & 6.76\% & 3.9570 & 65 & 7.8667 & 11.46\% & 6.5735 \\
        \midrule
        FPLM & 1 & 0.3563s & 150 & 2.2608 & 2.61\% & 1.6205 & 65 & 4.3559 & 5.11\% & 3.0819 \\
        ~& 2 & 0.1001s & 75 & 1.8124 & 2.02\% & 1.2610 & 65 & 4.4720 & 5.31\% & 3.1927 \\
        ~ & 3 & 0.0598s & 50 & 1.5319 & 1.65\% & 1.0337 & 65 & 4.5291 & 5.41\% & 3.2452 \\
        ~ & 5 & 0.0352s & 30 & 1.1849 & 1.21\% & 0.7585 & 65 & 4.6300 & 5.54\% & 3.3245 \\
        \midrule
\multicolumn{8}{l}{\underline{\hspace{-.075in}{\textit{Response variable: protein, non-functional covariate: fat}}}} \\
        FLM (B-spline) & 1 & 0.0078s & 150 & 1.5895 & 7.27\% & 1.2519 & 65 & 1.6474 & 7.53\% & 1.2918 \\
        ~& 2 & 0.0049s & 75 & 1.5464 & 7.10\% & 1.2245 & 65 & 1.6516 & 7.54\% & 1.2921 \\
        ~ & 3 & 0.0046s & 50 & 1.5014 & 6.91\% & 1.1930 & 65 & 1.6549 & 7.57\% & 1.2942 \\
        ~ & 5 & 0.0036s & 30 & 1.4203 & 6.55\% & 1.1325 & 65 & 1.6676 & 7.62\% & 1.3007 \\
        \midrule
        FLM (FPCA) & 1 & 0.0679s & 150 & 1.3717 & 6.47\% & 1.0538 & 65 & 1.4506 & 6.83\% & 1.1088 \\
        ~& 2 & 0.0575s & 75 & 1.3203 & 6.27\% & 1.0257 & 65 & 1.4374 & 6.76\% & 1.0968 \\
        ~ & 3 & 0.0321s & 50 & 1.2655 & 6.02\% & 0.9903 & 65 & 1.4272 & 6.72\% & 1.0893 \\
        ~ & 5 & 0.0291s & 30 & 1.1797 & 5.61\% & 0.9307 & 65 & 1.4200 & 6.68\% & 1.0810 \\
        \midrule
        FNPM & 1 & 0.2375s & 150 & 1.5264 & 7.37\% & 1.2227 & 65 & 2.3202 & 11.66\% & 1.8972 \\
        ~& 2 & 0.0817s & 75 & 1.4287 & 6.67\% & 1.1082 & 65 & 2.4228 & 12.43\% & 2.0100 \\
        ~ & 3 & 0.0605s & 50 & 1.6039 & 7.40\% & 1.2215 & 65 & 2.5539 & 13.29\% & 2.1386 \\
        ~ & 5 & 0.0378s & 30 & 1.8793 & 8.71\% & 1.4281 & 65 & 2.7117 & 14.31\% & 2.2919 \\
        \midrule
        FPLM & 1 & 0.3257s & 150 & 0.7640 & 2.97\% & 0.5051 & 65 & 1.3525 & 5.57\% & 0.9089 \\
        ~& 2 & 0.0818s & 75 & 0.6228 & 2.34\% & 0.3985 & 65 & 1.3894 & 5.85\% & 0.9503 \\
        ~ & 3 & 0.0467s & 50 & 0.5315 & 1.94\% & 0.3298 & 65 & 1.4045 & 5.98\% & 0.9670 \\
        ~ & 5 & 0.0367s & 30 & 0.4195 & 1.44\% & 0.2455 & 65 & 1.4523 & 6.25\% & 1.0078 \\
        \bottomrule
\end{longtable}
\end{center}

\vspace{-.5in}

Among the three models, the FPLM achieves the lowest errors for both the training and the testing data. As $K$ increases from 1 to 5, the execution time decreases markedly from 0.3061 seconds to 0.0378 seconds, whereas the prediction errors increase only slightly. The RMSFE increases from 1.8035 to 2.0126, the RFE from 12.89\% to 14.56\%, and the MAFE from 1.2856 to 1.4623. However, when $K = 5$, the discrepancy between training and testing errors becomes large. For example, the RMSE for the training data is 0.5904 compared with a RMSFE of 2.0126 for the testing data. This substantial gap indicates the presence of overfitting when the block sizes become too small.

For the FLM, the FPCA method yields lower prediction errors than the B-spline expansion method but requires a longer execution time. As $K$ increases from 1 to 5, the execution time decreases from 0.1155 seconds to 0.0195 seconds, while the prediction errors increase only modestly. For example, the RMSFE increases from 3.5582 to 3.5701.

Finally, the execution time for the FNPM is comparable to that of the FPLM, but the FNPM produces the largest estimation and prediction errors. Both the in-sample and out-of-sample errors are substantially higher than those of the FLM and FPLM.


Table~\ref{tbl:real_ECP} reports the mean ECP for all models. As $K$ increases from 1 to 5, the mean ECP for the training data changes only slightly. For the FLM and FNPM, the mean ECP for the testing data varies by less than 10\% across different values of $K$. In contrast, the FPLM exhibits a substantial decline in testing ECP as the number of blocks increases. For example, at the significance level $\alpha = 0.05$, the testing ECP drops from 86.32\% with one block to 55.14\% with five blocks. Consistent with the simulation study, increasing $K$ leads the FPLM to overfit the training data, producing much narrower prediction intervals.
As a consequence, a smaller proportion of testing observations fall within the intervals, resulting in a decrease in testing ECP.

\begin{center}
\tabcolsep 0.3in
\renewcommand{\arraystretch}{1.02}
\begin{longtable}{@{}lccccc@{}}
\caption{\small Mean empirical coverage probability (ECP) of the prediction intervals for all models under varying numbers of blocks. All results are averaged over 200 Monte Carlo resamplings.}
\label{tbl:real_ECP} \\
        \toprule
        ~ & ~ & \multicolumn{2}{c}{Training Data} & \multicolumn{2}{c}{Testing Data} \\
        \midrule
        Criterion & $K$ & $\alpha = 0.05$ & $\alpha = 0.2$ & $\alpha = 0.05$ & $\alpha = 0.2$ \\
        \midrule 
\endfirsthead
        \toprule
        ~ & ~ & \multicolumn{2}{c}{Training Data} & \multicolumn{2}{c}{Testing Data} \\
        \midrule
        Criterion & $K$ & $\alpha = 0.05$ & $\alpha = 0.2$ & $\alpha = 0.05$ & $\alpha = 0.2$ \\
        \midrule 
\endhead        
\multicolumn{6}{l}{\hspace{-.3in}{\underline{\textit{Response variable: fat, non-functional covariate: moisture}}}} \\
        FLM (B-spline) & 1 & 94.72\% & 80.10\% & 93.25\% & 77.41\% \\
        & 2 & 94.36\% & 79.58\% & 91.98\% & 75.22\% \\
        & 3 & 93.93\% & 79.12\% & 90.67\% & 73.58\% \\
        & 5 & 93.19\% & 78.90\% & 87.20\% & 70.57\% \\
        \midrule
        FLM (FPCA) & 1 & 94.68\% & 80.03\% & 92.84\% & 76.69\% \\
        & 2 & 94.14\% & 79.34\% & 91.68\% & 75.76\% \\
        & 3 & 93.73\% & 79.18\% & 90.71\% & 74.79\% \\
        & 5 & 93.00\% & 78.69\% & 88.53\% & 72.09\% \\
        \midrule
        FNPM & 1 & 94.68\% & 80.02\% & 79.38\% & 60.28\% \\
        & 2 & 94.09\% & 79.19\% & 74.05\% & 52.02\% \\
        & 3 & 93.05\% & 79.25\% & 74.31\% & 52.56\% \\
        & 5 & 91.86\% & 78.35\% & 77.22\% & 54.68\% \\
        \midrule 
        FPLM & 1 & 94.85\% & 80.06\% & 86.32\% & 59.35\% \\
        & 2 & 94.01\% & 79.66\% & 77.76\% & 48.96\% \\
        & 3 & 93.62\% & 79.97\% & 67.67\% & 42.37\% \\
        & 5 & 93.38\% & 80.47\% & 55.14\% & 32.12\% \\
        \midrule    
\multicolumn{6}{l}{\hspace{-.3in}{\underline{\textit{Response variable: moisture, non-functional covariate: protein}}}} \\
        FLM (B-spline) & 1 & 94.72\% & 80.06\% & 93.05\% & 77.95\% \\
        & 2 & 94.21\% & 79.64\% & 91.64\% & 76.43\% \\
        & 3 & 93.93\% & 79.23\% & 90.13\% & 75.18\% \\
        & 5 & 93.34\% & 79.11\% & 87.58\% & 71.54\% \\
        \midrule
        FLM (FPCA) & 1 & 94.73\% & 80.05\% & 92.82\% & 77.45\% \\
        & 2 & 94.19\% & 79.49\% & 91.50\% & 76.51\% \\
        & 3 & 93.78\% & 79.12\% & 90.18\% & 75.33\% \\
        & 5 & 92.88\% & 78.32\% & 87.93\% & 72.88\% \\
        \midrule
        FNPM & 1 & 94.71\% & 80.05\% & 78.60\% & 59.18\% \\
        & 2 & 94.18\% & 79.26\% & 72.92\% & 50.24\% \\
        & 3 & 93.20\% & 78.94\% & 73.15\% & 49.84\% \\
        & 5 & 91.94\% & 77.78\% & 76.70\% & 53.42\% \\
        \midrule 
        FPLM & 1 & 94.67\% & 80.04\% & 77.46\% & 55.53\% \\
        & 2 & 94.31\% & 79.79\% & 67.53\% & 45.74\% \\
        & 3 & 93.89\% & 80.00\% & 59.63\% & 39.16\% \\
        & 5 & 93.30\% & 79.79\% & 47.78\% & 29.13\% \\
        \midrule
\multicolumn{6}{l}{\hspace{-.3in}{\underline{\textit{Response variable: protein, non-functional covariate: fat}}}} \\
        FLM (B-spline) & 1 & 94.76\% & 80.05\% & 93.42\% & 78.35\% \\
        & 2 & 94.28\% & 79.61\% & 92.35\% & 77.22\% \\
        & 3 & 94.11\% & 79.65\% & 91.40\% & 75.78\% \\
        & 5 & 93.11\% & 79.42\% & 88.32\% & 72.93\% \\
        \midrule
        FLM (FPCA) & 1 & 94.71\% & 80.04\% & 93.41\% & 78.22\% \\
        & 2 & 94.50\% & 79.83\% & 93.04\% & 77.31\% \\
        & 3 & 94.18\% & 79.62\% & 92.08\% & 76.15\% \\
        & 5 & 93.48\% & 79.31\% & 89.90\% & 73.24\% \\
        \midrule
        FNPM & 1 & 94.74\% & 80.06\% & 80.40\% & 59.63\% \\
        & 2 & 94.14\% & 79.24\% & 73.70\% & 50.24\% \\
        & 3 & 93.04\% & 78.60\% & 74.40\% & 50.57\% \\
        & 5 & 91.88\% & 77.09\% & 77.66\% & 53.49\% \\
        \midrule 
        FPLM & 1 & 94.72\% & 80.03\% & 84.70\% & 60.21\% \\
        & 2 & 94.46\% & 79.90\% & 75.50\% & 49.41\% \\
        & 3 & 94.27\% & 80.26\% & 68.58\% & 42.18\% \\
        & 5 & 94.04\% & 80.67\% & 56.59\% & 31.94\% \\
        \bottomrule
\end{longtable}
\end{center}

\vspace{-.5in}

Table~\ref{tbl:real_IS} presents the mean IS for all models.
Consistent with the results in Table~\ref{tbl:real_ECP}, the IS for both the FLM and FNPM changes only slightly as $K$ increases.
More specifically, the IS for the FNPM increases, whereas the IS for the FLM decreases. This behaviour reflects the underlying estimation errors. For the FNPM, the prediction error increases slightly with larger $K$, leading to wider prediction intervals and higher IS values. However, for the FPLM, the IS exhibits substantial changes as $K$ increases to five. For example, at $\alpha = 0.05$, the IS for the training data decreases sharply from 6.5626 to 3.7501, while the IS for the testing data increases from 12.6356 to 27.5887. This pattern again reflects overfitting when the block size becomes too small. When $K = 5$, the estimation error for the training data is very small, producing narrow prediction intervals and thus a low IS. However, the corresponding ECP for the testing data is also low, meaning that many testing observations fall outside these overly narrow intervals. Consequently, the IS for the testing data becomes much larger.

\begin{center}
\tabcolsep 0.303in
\renewcommand{\arraystretch}{0.82}
\begin{longtable}{@{}lccccc@{}}
\caption{\small Mean interval score (IS) of the prediction intervals for all models under varying numbers of blocks. All results are averaged over 200 Monte Carlo replications.}
\label{tbl:real_IS} \\
\toprule
~ & ~ & \multicolumn{2}{c}{Training Data} & \multicolumn{2}{c}{Testing Data} \\
\midrule
Criterion & $K$ & $\alpha = 0.05$ & $\alpha = 0.2$ & $\alpha = 0.05$ & $\alpha = 0.2$ \\
\midrule 
\endfirsthead
\toprule
~ & ~ & \multicolumn{2}{c}{Training Data} & \multicolumn{2}{c}{Testing Data} \\
\midrule
Criterion & $K$ & $\alpha = 0.05$ & $\alpha = 0.2$ & $\alpha = 0.05$ & $\alpha = 0.2$ \\
\midrule 
\endhead        
\multicolumn{6}{l}{\hspace{-.3in}{\underline{\textit{Response variable: fat, non-functional covariate: moisture}}}} \\
FLM (B-spline)  & 1 & 22.0441 & 16.4233 & 24.4219 & 17.6411 \\
                & 2 & 21.4951 & 16.0297 & 24.7398 & 17.8333 \\
                & 3 & 20.8970 & 15.6031 & 25.4121 & 18.0406 \\
                & 5 & 19.9701 & 14.8478 & 27.4773 & 18.5540 \\
\midrule
FLM (FPCA)  & 1 & 16.3083 & 11.4966 & 18.9812 & 12.6204 \\
            & 2 & 15.7430 & 11.1535 & 19.6721 & 12.6725 \\
            & 3 & 15.0901 & 10.7836 & 20.4478 & 12.7205 \\
            & 5 & 13.9522 & 10.0791 & 22.1615 & 12.9191 \\
\midrule
FNPM        & 1 & 26.6415 & 19.6655 & 63.3199 & 35.6100 \\
            & 2 & 26.9487 & 19.0900 & 72.6947 & 39.5891 \\
            & 3 & 32.3489 & 21.8528 & 73.7530 & 41.3284 \\
            & 5 & 39.3029 & 26.1591 & 70.5169 & 42.6668 \\
\midrule
FPLM        & 1 & 6.5626 & 4.2027 & 12.6356 & 7.6570 \\
            & 2 & 5.5417 & 3.4592 & 16.4605 & 8.6691 \\
            & 3 & 4.6988 & 2.9016 & 20.8377 & 9.4773 \\
            & 5 & 3.7501 & 2.2545 & 27.5887 & 10.7842 \\
\midrule    
\multicolumn{6}{l}{\hspace{-.3in}{\underline{\textit{Response variable: moisture, non-functional covariate: protein}}}} \\
FLM (B-spline)  & 1 & 16.6094 & 12.8201 & 18.5330 & 13.6903 \\
                & 2 & 16.3121 & 12.5255 & 18.9859 & 13.7997 \\
                & 3 & 15.9286 & 12.2026 & 19.6644 & 13.9612 \\
                & 5 & 15.2902 & 11.6020 & 21.3095 & 14.3201 \\
\midrule
FLM (FPCA)      & 1 & 12.9184 & 9.5830 & 14.6998 & 10.3955 \\
                & 2 & 12.6243 & 9.3573 & 15.4068 & 10.4891 \\
                & 3 & 12.2062 & 9.0899 & 16.2291 & 10.6094 \\
                & 5 & 11.5409 & 8.5659 & 18.0676 & 10.8925 \\
\midrule
FNPM            & 1 & 19.5157 & 14.8766 & 45.4947 & 26.4862 \\
                & 2 & 19.4769 & 14.2343 & 52.7449 & 29.7252 \\
                & 3 & 23.5492 & 16.2415 & 53.8459 & 31.0859 \\
                & 5 & 29.2926 & 19.7284 & 51.4558 & 31.9695 \\
\midrule
FPLM            & 1 & 12.7646 & 8.1292 & 37.4739 & 18.1500 \\
                & 2 & 10.3283 & 6.6241 & 47.4894 & 20.5748 \\
                & 3 & 8.8720 & 5.6957 & 55.1092 & 22.1854 \\
                & 5 & 7.1097 & 4.4895 & 68.5854 & 24.7925 \\
\midrule
\multicolumn{6}{l}{\hspace{-.3in}{\underline{\textit{Response variable: protein, non-functional covariate: fat}}}} \\
FLM (B-spline)  & 1 & 7.4720 & 5.6147 & 8.0816 & 5.9077 \\
                & 2 & 7.2149 & 5.4572 & 8.2794 & 5.9333 \\
                & 3 & 6.9753 & 5.2937 & 8.4376 & 5.9593 \\
                & 5 & 6.6466 & 5.0082 & 9.0431 & 6.0563 \\
\midrule
FLM (FPCA)      & 1 & 7.1004 & 4.8131 & 8.1230 & 5.1821 \\
                & 2 & 6.6492 & 4.6053 & 8.0438 & 5.1286 \\
                & 3 & 6.2708 & 4.4156 & 8.0646 & 5.0833 \\
                & 5 & 5.7923 & 4.1331 & 8.2826 & 5.0933 \\
\midrule
FNPM            & 1 & 6.9003 & 5.3340 & 14.3843 & 8.9889 \\
                & 2 & 6.8773 & 5.0921 & 17.5235 & 10.1068 \\
                & 3 & 8.0607 & 5.7839 & 17.7427 & 10.5457 \\
                & 5 & 9.5969 & 6.7964 & 17.0190 & 10.8129 \\
\midrule
FPLM            & 1 & 4.7066 & 2.8086 & 10.6579 & 5.4566 \\
                & 2 & 3.8907 & 2.2947 & 13.1501 & 6.1707 \\
                & 3 & 3.3728 & 1.9688 & 15.0646 & 6.6155 \\
                & 5 & 2.7304 & 1.5531 & 19.0055 & 7.4875 \\
\bottomrule
\end{longtable}
\end{center}

\vspace{-.5in}

\FloatBarrier

\section{Conclusion and Future Perspectives}
\label{sec:conclusion}

This paper presents a unified distributed framework for point and interval estimation in three widely used scalar-on-function regression models, the functional linear model (FLM), the functional non-parametric model (FNPM), and the functional partial linear model (FPLM). The proposed approach allows each model to be fitted locally on distributed data blocks, with only aggregated intermediate results transmitted to a central server. This design preserves data privacy and offers substantial computational efficiency, making the framework appropriate for large-scale functional data and multi-institutional settings. 

Simulation results demonstrate that distributed estimation performs very effectively. For all three models, data partitioning leads to minimal loss of estimation or predictive accuracy while achieving large reductions in execution time. However, the FPLM is more sensitive to sample size. When blocks become too small, it tends to overfit the training data, producing large point estimation error, excessively narrow prediction intervals and reduced empirical coverage for new observations. The empirical analysis using the \textit{tecator} dataset reinforces these findings. Across all models, execution time decreases sharply as the number of blocks increases, while prediction errors remain largely stable. Among the models considered, the FPLM provides the best overall predictive accuracy, although it is also the most sensitive to excessive partitioning. 

This study can be extended in several directions, and we briefly mention two. First, we only consider scalar-on-function regression models, while it can be extended to function-on-scalar regression and function-on-function regression. Second, we show that there is a difference between the theoretical computation time and the empirical results. One may interested in solving such discrepancy.

\section*{Acknowledgement}

The authors are grateful for financial support from a Data Horizon project grant at Macquarie University.

\section*{CRediT Author Contributions}

\textbf{Peilun He:} Methodology, Software, Validation, Formal analysis, Investigation, Writing - Original Draft, Writing - Review \& Editing. 

\noindent \textbf{Nan Zou:} Conceptualization, Methodology, Formal analysis, Investigation, Writing - Original Draft, Writing - Review \& Editing, Supervision. 

\noindent \textbf{Han Lin Shang:} Conceptualization, Methodology, Software, Formal analysis, Investigation, Writing - Original Draft, Writing - Review \& Editing, Visualization, Supervision, Funding acquisition. 

\section*{Disclosure statement}
The authors report there are no competing interests to declare.

\section*{Data Availability}

The codes for the Monte Carlo simulation and empirical analysis are available on GitHub at: \url{https://github.com/peilun-he/Distributed-Functional-Regression}.

\newpage
\bibliographystyle{apalike}
\bibliography{DL_FLM.bib}

\end{document}